\documentclass[acmsmall,manuscript,screen]{acmart}
\AtBeginDocument{%
  }

\setcopyright{acmcopyright}
\copyrightyear{2018}
\acmYear{2018}
\acmDOI{XXXXXXX.XXXXXXX}

\usepackage{xspace}
\usepackage[inline]{enumitem}
\usepackage{fancyvrb}
\usepackage{xcolor}
\usepackage{soul}
\usepackage{tabularx}
\usepackage{algorithm}
\usepackage{algpseudocode}

\VerbatimFootnotes

\newcommand{\urlarr}{\url{https://www.clpset.unipr.it/SETLOG/APPLICATIONS/array.zip}}

\newcommand{\setlog}{$\{log\}$\xspace}

\newcommand{\defs}{\triangleq}
\DeclareMathSymbol{\dres}{\mathbin}{AMSa}{"43}
\DeclareMathSymbol{\rres}{\mathbin}{AMSa}{"42}
\def \ndres     {\mathbin{\rlap{\hspace{.2ex}\hbox{$-$}}{\dres}}}
\def \nrres     {\mathbin{\rlap{\raise.05ex\hbox{$-$}}{\rres}}}
\def \lover		  {\mathbin{{\dres} \llap{+\!}\;}}

\newcommand{\num}{\mathbb{Z}}
\newcommand{\nat}{\mathbb{N}}

\newcommand{\Pfun}{\mathit{pfun}}
\newcommand{\Dom}{\mathit{dom}}
\newcommand{\Card}[1]{\vert #1 \vert}
\newcommand{\true}{\mathit{true}}
\newcommand{\false}{\mathit{false}}

\newcommand{\plus}{\mathbin{\scriptstyle\sqcup}}
\newcommand{\e}{\emptyset}
\newcommand{\w}{\{\cdot \plus \cdot\}}
\renewcommand{\i}{[\cdot,\cdot]}
\renewcommand{\Cup}{un}
\newcommand{\disj}{\parallel}

\renewcommand{\Cap}{\mathit{inters}}
\newcommand{\Diff}{\mathit{diff}}

\newcommand{\Dres}{\mathit{dres}}

\newcommand{\Dares}{\mathit{dares}}

\newcommand{\Get}{\mathit{get}}
\newcommand{\Upd}{\mathit{upd}}
\newcommand{\Ifun}{\mathit{ipfun}}
\newcommand{\Pair}{\mathit{pair}}
\newcommand{\Npair}{\mathit{npair}}
\newcommand{\Rel}{\mathit{rel}}
\newcommand{\Arr}{\mathit{arr}}
\newcommand{\Ncup}{\mathit{nun}}

\newcommand{\QRI}{\textsf{QA}}
\newcommand{\QA}{\QRI}
\newcommand{\LQRI}{\mathcal{L}_{\QRI}}
\newcommand{\LQA}{\LQRI}
\newcommand{\SQA}{\mathcal{S}_{\QA}}

\newcommand{\INT}{[\,]}
\newcommand{\LINT}{\mathcal{L}_{\INT}}
\newcommand{\SATINT}{\mathcal{SAT}_{\INT}}
\newcommand{\RQ}{\mathcal{RQ}}
\newcommand{\LRQ}{\mathcal{L}_\RQ}
\newcommand{\SATRQ}{\mathit{SAT}_\RQ}
\newcommand{\TU}{\mathcal{U}}

\newcommand{\STEPCARD}{\mathsf{solve\_size}}

\newcommand{\Var}{\mathcal{V}}
\newcommand{\Set}{\mathsf{S}}
\newcommand{\Int}{\mathsf{Z}}
\newcommand{\OPair}{\mathsf{P}}   
\newcommand{\Ur}{\mathsf{U}}
\newcommand{\FSet}{\mathcal{F}_\Set}
\newcommand{\FInt}{\mathcal{F}_\Int}
\newcommand{\FPair}{\mathcal{F}_\OPair}   
\newcommand{\FUr}{\mathcal{F}_\Ur}
\newcommand{\PiSet}{\Pi_\mathsf{S}}
\newcommand{\Size}{size}
\newcommand{\PiInt}{\Pi_\Int}
\newcommand{\sSet}{\mathsf{Set}}
\newcommand{\sInt}{\mathsf{Int}}
\newcommand{\sUr}{\mathsf{Ur}}
\newcommand{\sPair}{\mathsf{OPair}}    
\newcommand{\TQRI}{\mathcal{T}_{\QRI}}
\newcommand{\TInt}{\mathcal{T}_\Int}
\newcommand{\TUr}{\mathcal{T}_\Ur}
\newcommand{\frm}{\mathfrak{F}}
\newcommand{\FQRI}{\frm_\QRI}

\newcommand{\iS}{\mathcal{R}}
\newcommand{\iF}[1]{(#1)^\iS}

\newcommand{\why}[1]{\tag*{{\footnotesize [by #1]}}}
\newcommand{\by}[1]{{\footnotesize [by #1]}}

\newcommand{\q}{\text{\normalfont\'{}}}
\newcommand{\qr}{\text{\normalfont\'{}}\hspace{5pt}}
\newcommand{\ql}{\hspace{5pt}\text{\normalfont\'{}}}
\newcommand{\lfun}{\longrightarrow}

\newtheorem{remark}{Remark}

\newif\ifcomments
\commentstrue 

\newcommand{\maxi}[1]{\ifcomments\sethlcolor{green}\hl{M: #1}\fi}

\begin{document}

\title{Encoding and Reasoning about Arrays in Constraint Logic Programming with Sets}

\author{Maximiliano Cristi\'a}
\email{cristia@cifasis-conicet.gov.ar}
\affiliation{%
  \institution{Universidad Nacional de Rosario and CIFASIS}
  \city{Rosario}
  \country{Argentina}
}


\author{Gianfranco Rossi}
\email{gianfranco.rossi@unipr.it}
\affiliation{%
  \institution{Universit\`a di Parma}
  \city{Parma}
  \country{Italy}}

\begin{abstract}
We encode arrays as functions which, in turn, are encoded as sets of ordered pairs. The set cardinality of each of these functions coincides with the length of the array it is representing. Then we define a fragment of set theory that is used to give the specifications of a non-trivial class of programs with arrays. In this way, array reasoning becomes set reasoning. Furthermore, a decision procedure for this fragment is also provided and implemented as part of the \setlog (read `setlog') tool. \setlog is a constraint logic programming language and satisfiability solver where sets and binary relations are first-class citizens. The tool already implements a few decision procedures for different fragments of set theory. In this way, arrays are seamlessly integrated into \setlog thus allowing users to reason about sets,  functions and arrays all in the same language and with the same solver. The decision procedure presented in this paper is an extension of decision procedures defined in earlier works not supporting arrays.
\end{abstract}



\received{20 February 2007}
\received[revised]{12 March 2009}
\received[accepted]{5 June 2009}

\maketitle

\RecustomVerbatimEnvironment{Verbatim}{Verbatim}{xleftmargin=8mm,fontfamily=courier}

\RecustomVerbatimCommand{\Verb}{Verb}%
  {fontfamily=courier}

\section{\label{intro}INTRODUCTION}
Consider functions as sets of ordered pairs and let $\Pfun(F)$ be a constraint stating that a set of ordered pairs $F$ is a function.\footnote{$\Pfun$ stands for \emph{partial function}, thus emphasizing the fact that in this context functions are sets of ordered pairs rather than primitive objects as in type theory and functional programming.}
Hence, the fact that a set of ordered pairs $A$ is an array of length $m \in \nat$ (i.e., a \emph{finite} array) can be expressed in terms of set theory as follows\footnote{Considering that array indexes begin at 1.}.
\begin{equation}\label{eq:arr}
arr(A,m) \iff \Pfun(A) \land \Card{A} = m \land (\forall (x,y) \in A: 1 \leq x \leq m)
\end{equation}
A predicate of the form $\forall x \in S : \phi$, for some set $S$ and formula $\phi$, is called restricted universal quantifier (RUQ) and is interpreted as $\forall x (x \in S \implies \phi)$.

Now consider a program checking whether or not array \Verb+A+ of length \Verb+m+ is sorted.
\begin{Verbatim}
bool sorted(A,m) {
  k := 1;
  while (k < m && A[k] =< A[k+1]) k := k + 1;
  return k = m;
}
\end{Verbatim}
If we want to formally verify \Verb+sorted+ we can encode array \Verb+A+ as a function $A$ satisfying $arr(A,m)$.\footnote{We use \Verb+verbatim font+ for programs and \textit{math font} for specifications.} In this way the specification of \Verb+sorted+ can be stated in terms of its pre- and post-condition as follows.
\begin{align*}
& \textsc{Pre: } arr(A,m) \\
& \textsc{Post: } \text{return} = true \iff \forall (i_1,x_1), (i_2,x_2) \in A : i_1 \leq i_2 \implies x_1 \leq x_2
\end{align*}

The post-condition of \Verb+sorted+ can be simplified if we introduce the following predicate.
\begin{equation}\label{sorted}
\mathit{sorted}(A) \iff \forall (i_1,x_1), (i_2,x_2) \in A : i_1 \leq i_2 \implies x_1 \leq x_2
\end{equation}
Furthermore, a good loop invariant to verify \Verb+sorted+ is that the left part of \Verb+A+ is sorted. In terms of set theory, the left part of an array can be defined by means of the so-called \emph{domain restriction} operator, noted $\dres$ \cite{Spivey00}.
\begin{equation}\label{eq:dres}
I \dres A = \{(x,y) \mid (x,y) \in A \land x \in I\}
\end{equation}
Note that if $A$ is an array of length $m$ and $k \leq m$ then $[1,k] \dres A$ is also an (smaller) array\footnote{$[k,m]$ denotes the closed integer interval from $k$ to $m$.}. Now the loop invariant can be given as follows.
\begin{equation}\label{inv_sorted}
\textsc{Inv: } \mathit{sorted}([1,k] \dres A)
\end{equation}
It is easy to see that domain restriction can be put in terms of basic set operators and RUQ ($N$ is an existential variable)\footnote{This is one of the many ways in which domain restriction can be written.}.
\begin{equation}
B = I \dres A \iff
  A = B \cup N \land (\forall (x,y) \in B: x \in I)
  \land (\forall (x,y) \in N : x \notin I)
\end{equation}
Domain restriction is a key operation when specifying loop invariants of programs with arrays as most often loops process arrays from left to right thereby establishing the desired property on the left part of the array as the loop progresses.

In summary, an approach to reason about programs with arrays needs to be able to reason about functions as set of ordered pairs, set cardinality, integer intervals and RUQ.  RUQ are important because in the specifications of these programs, in many cases, it will be necessary to state properties about all the components of the arrays.
Moreover, to \emph{automatically} reason about these programs, it would be necessary to find decision procedures for the theory outlined above.

In a previous work \cite{DBLP:journals/tocl/CristiaR24} we have shown a decision procedure for the Boolean algebra of finite sets extended with set cardinality, integer intervals and linear integer arithmetic. We have also provided a decision procedure for a certain fragment of a theory including RUQ \cite{DBLP:journals/jar/CristiaR24}.
In this paper we provide a decision procedure for a combination of these theories thus allowing us to automatically reason about a class of programs with arrays---including, for instance, the one just mentioned.
In this way, reasoning about arrays becomes reasoning about sets, set cardinality, functions (as set of ordered pairs) and integer intervals. Furthermore, many standard set and relational operators (e.g., domain restriction) can be used to give the specifications and loop invariants of these programs.
This should be particularly helpful for users of set-based notations such as B \cite{Abrial00,Abrial:2010:MES:1855020} and Z \cite{Spivey00} because in this way they can include arrays in their specifications and reason about them inside the same theoretical framework.
Moreover, all the theory and practice already developed by us for other fragments of set theory (e.g., \cite{Dovier00,DBLP:journals/jar/CristiaR20}) is used here to address arrays. 

The language and the decision procedure presented in this paper aim at reaching the results presented by Bradley, Manna, and Sipma \cite{DBLP:conf/vmcai/BradleyMS06}, perhaps one of the most  expressive existing theories of infinite arrays.
In the end, our results are slightly more powerful than those of Bradley et al.\footnote{From now one we will use Bradley as a shorthand for the article by Bradley et al. \cite{DBLP:conf/vmcai/BradleyMS06}.} because it is possible to reason about more array formulas and it is possible to reason about functions, cardinality and intervals in general.
Since that work, others (e.g., \cite{DBLP:conf/fossacs/HabermehlIV08}) have provided decision procedures for more advanced fragments of the theory of arrays. We believe that showing that at least Bradley's results can be reached from set theory is enough for a first work since, as we just said, this seamlessly integrates arrays into a framework where sets are first-class entities.

The new decision procedure is implemented as part of \setlog (read `setlog') \cite{setlog}, a constraint logic programming (CLP) language and satisfiability solver where sets and binary relations are first-class entities.
The tool is the practical result of our theoretical work carried out over the years on CLP with sets \cite{DBLP:journals/jlp/DovierOPR96,Dovier00}.
Broadly speaking, \setlog works as an extension of Prolog to logic set programming \cite{DBLP:books/daglib/0067831,DBLP:series/mcs/CantoneOP01} thus enabling applications such as formal verification of and test case generation from set formulas, as is shown in this paper.
Several in-depth empirical evaluations provide evidence that \setlog is able to solve non-trivial problems \cite{DBLP:journals/jar/CristiaR20,DBLP:conf/RelMiCS/CristiaR18,DBLP:journals/jar/CristiaR21a,CristiaRossiSEFM13}; in particular as an automated verifier of security properties \cite{DBLP:journals/jar/CristiaR21,DBLP:journals/jar/CristiaR21b,DBLP:journals/jar/CristiaLL23,DBLP:conf/csfw/CapozuccaCHK24}.
\setlog now offers not only programming with and automated reasoning about sets but also a set-based state machine specification language, a dedicated environment for executing those state machines, a verification condition generator and automated test case generation \cite{CRISTIA_CAPOZUCCA_ROSSI_2026}.
Therefore, by providing automated reasoning for programs with arrays \setlog now offers a more comprehensive environment for formal verification.

\paragraph{Contributions of the paper}
The contribution of the paper is both theoretical and practical.
\begin{enumerate}
\item From the theoretical side, we show that the decision problem for a theory of arrays can be approached from set theory by profiting from existing decidability results about set theory.
\item From a practical stand-point:
\begin{enumerate*}
\item The implementation of the decision procedure as part of the \setlog tool provides an environment where users can reason about arrays, sets, functions, etc. with the same notation and operators;
\item Two applications of \setlog as a solver for arrays problems are shown, namely, formal verification and model-based test case generation.
\end{enumerate*}
\end{enumerate}

\paragraph{Structure of the paper}
The paper is organized as follows.
Section \ref{language} consists of a formal presentation of our language, $\LQRI$.
In Section \ref{expr} we show the expressive power of $\LQA$ and compare it with Bradley's language.
A solver for $\LQA$ is described in Section \ref{solver} where we include the definition of irreducible formula.
In Section \ref{proofs} we prove that the solver is a decision procedure for $\LQA$ formulas by taking advantage of previous results.
The implementation of $\LQA$ as part of the \setlog tool is introduced in Section \ref{setlog}.
Afterwards, in Section \ref{applications}, two applications of \setlog as a solver for array problems are showcased. Specifically, we show how \setlog can be used in formal verification and as a test case generator of programs with arrays.
In Section \ref{relwork} we discuss our results in comparison with other approaches to the same problem.
Our conclusions are given in Section \ref{concl}.

\section{\label{language}$\LQRI$, A QUANTIFIED LANGUAGE FOR ARRAYS ENCODED IN SET THEORY}

In this section we give a formal account of the set-based language $\LQRI$ that allows the formalization of a class of array properties.
$\LQA$ is a combination between the languages known as $\LINT$ (read `l-interval') \cite{DBLP:journals/tocl/CristiaR24} and $\LRQ$ \cite{DBLP:journals/jar/CristiaR24}, plus some additions.
In this way, arrays become functions which in turn are encoded as binary relations---i.e., sets of ordered pairs---thus turning array formulas into set theoretic formulas.
$\LINT$ is a multi-sorted first-order predicate language which allows the specification of problems involving finite extensional sets, finite integer intervals, set cardinality, the Boolean operators of set theory (basically union, intersection and difference), and linear integer arithmetic (LIA).
$\LRQ$ is a first-order predicate language with terms designating sets and terms designating ur-elements\footnote{Ur-elements (also known as atoms or individuals) are objects which contain no elements but are distinct from the empty set. ``Ur'' is a German prefix meaning ``primitive'' or ``original''}.
Terms designating ur-elements are provided by an external quantifier-free, first-order, decidable theory $\TU$.
$\LRQ$ is configured with $\TU$ thus extending the latter with restricted quantifiers of the form $(\forall c \in S : \phi)$ (RUQ) and $(\exists c \in S : \phi)$ (REQ) where $c$ is either a variable or an ordered pair of distinct variables, $S$ is a finite extensional set and $\phi$ is a $\TU$ formula.
In $\LQRI$ REQ will not be considered.

\subsection{Syntax}

The syntax of the language is defined primarily by giving the signature upon which terms and formulas are built.

\begin{definition}[Signature]\label{signature}
The signature $\Sigma_{\QRI}$ of $\LQRI$ is a triple $\langle
\mathcal{F},\Pi,\Var\rangle$ where:
\begin{itemize}
\item $\mathcal{F}$ is the set of function symbols along with their sorts, partitioned as
$\mathcal{F} \defs \FSet \cup \FInt \cup \FPair \cup \FUr$, where
$\FSet \defs \{\e,\w,\i\}$, 
$\FInt \defs \{0,-1,1,-2,2,\dots\} \cup \{+,-,*\}$,
$\FPair \defs \{(\cdot,\cdot)\}$, and
$\FUr$ is a possibly empty set of uninterpreted constant and function symbols.
\item $\Pi$ is the set of predicate symbols along with their sorts, partitioned as
$\Pi \defs \Pi_{=} \cup \PiSet \cup \Pi_{\Size} \cup \PiInt \cup \Pi_{\OPair}$, where 
$\Pi_{=} \defs  \{=,\neq\}$, 
$\PiSet \defs \{\in,\notin,\Cup,\disj\}$, 
$\Pi_{\Size} \defs \{\Size\}$,
$\PiInt \defs \{<\}$, and
$\Pi_{\OPair} \defs \{\Pair,\Npair\}$.
\item $\Var$ is a denumerable set of variables partitioned as
$\Var \defs \Var_\Set \cup \Var_\Int \cup \Var_\OPair \cup \Var_\Ur$.
\end{itemize}
\end{definition}

Intuitively, 
$\e$ represents the empty set; 
$\{x \plus E\}$ represents the set $\{x\} \cup E$; 
$[m,n]$ represents the set $\{p \in \num | m \leq p \leq n\}$; 
$(\cdot,\cdot)$ will be used to represent ordered pairs; and
$\Var_\Set$, $\Var_\Int, \Var_\OPair$ and $\Var_\Ur$ represent sets of variables ranging over sets, integers, ordered pairs and ur-elements, respectively.
Along the same lines, $\in$, $\notin$ and $<$ have their standard meaning.
On the other hand, $\Cup(E,F,G)$ is interpreted as $G = E \cup F$, $E \disj F$ as $E \cap F = \e$ and $\Size(E,n)$ as $\Card{E} = n$.
Finally, $\Pair(x)$ holds only if $x$ unifies with an ordered pair, i.e., with a term of the form $(\cdot,\cdot)$; $\Npair$ has the opposite meaning.

$\LQRI$ is a many-sorted language whose sorts are $\sSet$, $\sInt$, $\sPair$ and $\sUr$, representing sets, integers, ordered pairs and ur-elements, respectively.
The sort of $\e$ is $\sSet$ and the sort of $\{x \plus E\}$ is $\sSet$ provided $E$ is of sort $\sSet$ ($x$ can be of any sort).
The sort of $(x,y)$ is $\sPair$.
The sort of $f \in \FUr$ is $\sUr$ provided all its arguments are of sort $\sUr$ as well. 
Clearly,  $v \in \Var_\Set$ is of sort $\sSet$, $v \in \Var_\Int$ is of sort $\sInt$, $v \in \Var_\OPair$ is of sort $\sPair$ and $v \in \Var_\Ur$ is of sort $\sUr$.
Concerning predicate symbols,
$=$ and $\neq$ expect two terms of any sort; 
in $x \in E$ and $x \notin E$, $x$ can be of any sort, while $E$ must be of sort $\sSet$;
in $\Cup(E,F,G)$ and $E \disj G$ all arguments must be of sort $\sSet$;
in $\Size(E,n)$, $E$ must be of sort $\sSet$ whereas $n$ must be of sort $\sInt$;
$<$ expects two arguments of sort $\sInt$; and
$\Pair$ and $\Npair$ expect a term of any sort.
Note that arguments of $=$ and $\neq$ can be of any of the four considered sorts. We do not have distinct symbols for different sorts, but the interpretation of $=$ and $\neq$ (see Section \ref{semantics}) depends on the sorts of their arguments.

The set of admissible (i.e., well-sorted) $\LQRI$ terms is defined as follows.

\begin{definition}[$\QRI$-terms]\label{terms}
The set of \emph{$\QRI$-terms}, denoted by $\TQRI$, is the minimal subset of
the set of $\Sigma_{\QRI}$-terms generated by the following grammar complying
with the sorts as given above:
\begin{flalign*}
C &::= 0 \mid {-1} \mid 1 \mid {-2} \mid 2 \mid \dots \\
\TInt^0 &::=
  C 
  \mid \Var_\Int \\
\TInt &::=
  \TInt^0
  \mid C * \Var_\Int 
  \mid \Var_\Int * C \mid \TInt + \TInt \mid \TInt - \TInt \\
OP & ::=
  \Var_\OPair
  \mid \q(\qr \mathcal{T}_{\Int\Ur} \ql,\qr \mathcal{T}_{\Int\Ur} \ql)\q \\
\mathit{Set} & ::=
   \q\e\qr
   \mid \Var_S
   \mid \q\{\qr \TQRI \ql\plus\qr \mathit{Set} \ql\}\q
   \mid \q[\qr \TInt \ql,\qr \TInt \ql]\q \\
\TQRI & ::=
  OP \mid \TInt \mid \TUr \mid \mathit{Set} 
\end{flalign*}
where $\TUr$ represent any term of sort $\sUr$ and $\mathcal{T}_{\Int\Ur}$ represents any term in $\TInt \cup \TUr$. 
\end{definition}

As can be seen, concerning the sort $\sInt$ the grammar admits only linear terms.

In the following we will say that a term of sort $\sSet$ is a \emph{set term} or just a \emph{set} and likewise a term of sort $\sInt$ is an \emph{integer term} or just an \emph{integer}. 
In particular, terms of the form $\w$ are called \emph{extensional set terms} or just \emph{extensional sets}, and terms of the form $\i$ are called \emph{integer intervals} or just \emph{intervals}. 
The arguments of intervals are called \emph{left} and \emph{right limits}, respectively.
It is important to remark that interval limits can be linear integer  terms including variables.

\begin{remark}[Simplified notation for extensional sets]
Hereafter, we will use the following notation for extensional sets:
$\{t_1,t_2,\dots,t_n \plus t\}$ is a shorthand for $\{t_1 \plus \{t_2 \,\plus\,\cdots \{ t_n \plus t\}\cdots\}\}$, while 
$\{t_1,t_2,\dots,t_n\}$ is a shorthand for $\{t_1,t_2,\dots,t_n \plus \e\}$.
In $\{t_1,t_2,\dots,t_n \plus t\}$ we say $t$ is the \emph{set part}.
\end{remark}

The atomic predicates of $\LQRI$ are $\true$, $\false$, and the atoms formed by applying the predicate symbols in $\Pi$ to $\QRI$-terms of the right sorts as described above. 
Atoms built from predicate symbols in $\Pi$ whose arguments are of the right sort are called $\Pi$-constraints .
In this way, the $\QA$-formulas are defined as follows.

\begin{definition}[$\QRI$-formulas]\label{formulas}
The set of $\QRI$-formulas, denoted by $\FQRI$, is given by the following
grammar under the restrictions indicated below:
\begin{align*}
CT & ::=
  \Var_{\Int\Ur}
  \mid \q(\qr \Var_{\Int\Ur} \ql,\qr \Var_{\Int\Ur} \ql)\q \\
\mathit{QFV} &::=
  CT = CT
  \mid CT \neq CT
  \mid \Var_\Int < \Var_\Int \\
\mathit{QFT} &::=
  \mathcal{T}_{\Int\Ur} = \mathcal{T}_{\Int\Ur}
  \mid \mathcal{T}_{\Int\Ur} \neq \mathcal{T}_{\Int\Ur}
  \mid \TInt < \TInt
  \mid \Pair(OP) \\
\mathit{QF} &:: =
  \mathit{QFV}
  \mid \mathit{QFT}
  \mid \mathit{QF} \land \mathit{QF}
  \mid \mathit{QF} \lor \mathit{QF} \\
\mathit{RUQ} &::= 
  (\forall\; CT \in \mathit{Set}: \mathit{QF})
  \mid (\forall\; CT \in \mathit{Set}: \mathit{RUQ}) \\
\FQRI &::=
  \true \mid \false \mid \Pi 
  \mid \mathit{RUQ}
  \mid \FQRI \land \FQRI
  \mid \FQRI \lor \FQRI
\end{align*}
where:
$\Var_{\Int\Ur}$ represents any variable in $\Var_\Int \cup \Var_\Ur$;
if $(x,y)$ is an element of $CT$, then $x$ and $y$ must be distinct variables;
if $t$ is term of sort $\sInt$ in a $QFT$ atom and $t \notin \TInt^0$, then no quantified variable is admitted in $t$;
$\mathcal{T}_{\Int\Ur}$, $OP$, $\mathit{Set}$ and $\TInt$ are borrowed from Definition \ref{terms},  and $\Pi$ represents any $\Pi$-constraint.
\end{definition}

In $(\forall\; CT \in \mathit{Set}: \phi)$ we say that $CT$ is the \emph{quantified term} and $\mathit{Set}$ is the \emph{quantification domain} or just \emph{domain}. Note that the language admits variables and ordered pairs (formed only by distinct variables) as quantified terms. If the quantified term is a variable we say it is a quantified variable.
The RUQ resulting from the last case of rule $\mathit{RUQ}$ are said to be \emph{nested RUQ}.
Tables \ref{t:op} and \ref{t:bradley} show some examples of $\QRI$-formulas.

\begin{example}[$\FQRI$-formulas]
The following is a $\FQRI$ formula:
\[
\forall (x,y) \in R: k+1 < x \land x < m+5 \land 2*k < y \land y < 3*m
\]
However, the following is not a $\FQRI$ formula:
\[
\forall (x,y) \in R: k < x+1 \land x+1 < m \land k < 2*y \land 2*y < m
\]
because $x+1$ and $2*y$ are terms of sort $\sInt$ not belonging to $\TInt^0$ while $x$ and $y$ are quantified variables.
\end{example}

\begin{remark}
Bradley defines the set of array formulas for which they provide a decision procedure as a subset of the formulas of the form $\exists^*\forall^*_\num\,\phi$ where $\forall^*_\num$ means quantification over array indexes and $\phi$ is a quantifier-free formula of a particular form. Instead in $\LQA$ formulas are given in Skolem normal form as is usually the case in CLP and logic programming.
So, for instance, in $(\forall x \in R : \Pair(x))$ variable $R$ is implicitly existentially quantified.
\end{remark}

\begin{remark}\label{lint-lqa}
Definitions \ref{signature} and \ref{terms} are basically those of $\LINT$ \cite{DBLP:journals/tocl/CristiaR24} with the explicit addition of ordered pairs. $\LINT$ admits a set of uninterpreted function symbols but no particular one was included. Now we fix $(\cdot,\cdot)$ as a required symbol.

On the contrary, Definition \ref{formulas} is borrowed from $\LRQ$ \cite{DBLP:journals/jar/CristiaR24} with a couple of modifications. First, REQ have been eliminated.
Second, the quantifier-free formulas inside RUQ (i.e., the QF-formulas of Definition 2.3) are fixed to be propositions of atomic predicates formed with symbols in $\{=,\neq,<,\Pair\}$.

Furthermore, if $\mathit{RUQ}$ is removed from the definition of $\FQRI$, the class of formulas so produced is that of $\LINT$.
\end{remark}

\begin{remark}[Naming conventions]
We will use the following naming conventions: 
$R,S,T$ will be used for sets of ordered pairs;
$E, F, G, H$ will be used for any sets excluding integer intervals;
$i, j, k, m, n$ will denote integers;
$a, b, c, d$ will denote terms of sort $\sUr$; and 
$t, x, y, z$ will be used for terms of any of the four sorts.
\end{remark}

\begin{remark}[Other logical connectives in \textit{QF}]
It is possible to admit other logical connectives in $\mathit{QF}$ as long as the resulting formula is a $\mathit{QF}$-formula. For instance, $x = y \implies a = b$ is equivalent to $x \neq y \lor a = b$, which is a $\mathit{QF}$-formula.
\end{remark}

\begin{remark}[Simplified notation for RUQ]
Instead of writing $(\forall x \in E: (\forall y \in F: \phi))$ we will write $(\forall x \in E, y \in F: \phi)$. Likewise, instead of $(\forall x \in E, y \in E: \phi)$ we will write $(\forall x,y \in E: \phi)$
\end{remark}

\subsection{\label{semantics}Semantics}
Sorts and symbols in $\Sigma_{\QRI}$ are interpreted according to the interpretation structure $\iS \defs \langle D,\iF{\cdot}\rangle$, where $D$ and $\iF{\cdot}$ are defined as follows.

\begin{definition} [Interpretation domain] \label{def:int_dom}
The interpretation domain $D$ is partitioned as $D \defs D_\sSet \cup D_\sInt \cup D_\sUr$ where:
\begin{itemize}
\item $D_\sSet$ is the set of all hereditarily finite hybrid
sets built from elements in $D_\sInt \cup D_\sUr$.
Hereditarily finite sets are those sets that admit (hereditarily finite) sets as their elements, that is sets of sets.
\item $D_\sInt$ is the set of integer numbers, $\mathbb{Z}$.
\item $D_\sUr$ (i.e., the set of ur-elements) is the set of all uniterpreted ground terms of $\LQRI$ (including those built from $(\cdot,\cdot)$).

\end{itemize}
\end{definition}

\begin{definition} [Interpretation function] \label{app:def:int_funct}
The interpretation function $\iF{\cdot}$ for well-sorted terms and atoms of $\LQRI$ is defined as follows:
\begin{itemize}
\item Each sort $\mathsf{X} \in \{\sSet,\sInt\}$ is mapped to the domain $D_\mathsf{X}$, while sorts $\sUr$ and $\sPair$ are mapped to $D_\sUr$.
\item Terms built from constant and function symbols in $\mathcal{F}_\Set$ are mapped to objects in $D_\sSet$ as follows:
  \begin{itemize}
  \item $\e$ is interpreted as the empty set
  \item $\{ x \plus E \}$ is interpreted as the set $\{x^\iS\} \cup E^\iS$
  \item $[k,m]$ is interpreted as the set $\{p \in \num \mid k^\iS \leq p \leq m^\iS\}$ (note that $k^\iS$ and $m^\iS$ are objects in $D_\sInt$)\footnote{Note that integer intervals in $\LQRI$ denote always finite sets given that their limits can assume only integer values.}
  \end{itemize}

\item Terms built from constant and function symbols in $\mathcal{F}_\Int$ are mapped to objects and operations in $D_\sInt$ as follows:
\begin{itemize}
\item Each element in \{0,-1,1,-2,2,\dots\} is interpreted as the
corresponding integer number in $D_\sInt$
\item $i + j$, $i - j$, $i * j$ are interpreted as $i^\iS + j^\iS$, $i^\iS - j^\iS$,  $i^\iS * j^\iS$, respectively
\end{itemize}

\item Terms built from constant and function symbols in $\mathcal{F}_\Ur \cup \mathcal{F}_\OPair$ are mapped to themselves in $D_\sUr$

\item The predicate symbols in $\Pi_{=}$ are interpreted as follows:
  \begin{itemize}
   \item $x = y$, where $x$ and $y$ have the same sort $\mathsf{X}$,
   $\mathsf{X} \in \{\sSet,\sInt,\sPair,\sUr\}$, is interpreted  
   as the identity between $x^\iS$ and $y^\iS$ in $D_\mathsf{X}$
   \item $x \neq y$ is interpreted as $\lnot x = y$
  \end{itemize}

\item The predicate symbols in $\PiSet \cup \Pi_{\Size}$ are mapped to set-theoretical operations as follows:
  \begin{itemize}

   \item $x \in E$ is interpreted as $x^\iS \in E^\iS$ (note that $x^\iS$ is an object in $D_\sInt \cup D_\sUr$)
   \item $x \notin E$ is
   interpreted as $\lnot x \in E$
   \item $\Cup(E,F,G)$ is interpreted as $G^\iS = E^\iS \cup F^\iS$
   \item $E \disj F$ is interpreted as $E^\iS \cap F^\iS = \emptyset$
   \item $\Size(E,k)$ is interpreted as $\Card{E^\iS} = k^\iS$ (note that $k^\iS$ is an object in $D_\sInt$)
  \end{itemize}

\item The predicate symbol $<$ is mapped to the corresponding relation between integers as follows: $i < j$ is interpreted as $i^\iS < j^\iS$
  
  \item The predicate symbols in $\Pi_{\OPair}$ are mapped to equality in $D_\sUr$ (i.e., syntactic equality) as follows: $\Pair(x)$ is interpreted as $\exists a,b(x^\iS = (a,b))$, while $\Npair(x)$ is interpreted as $\lnot\Pair(x)$
  
  \item The RUQ $(\forall x \in E: \phi)$ is interpreted as
   $(\forall x(x \in E^\iS \implies \phi^\iS))$
\end{itemize}
\end{definition}

Set (resp., integer) operators and predicates between elements of the domain $D$ are interpreted as usual in set (resp., integer) theory. All other operators and connectives are interpreted according to the standard rules of first-order predicate logic.

In particular, observe that equality between two set terms is mapped by $\iS$ to equality in $D_\sSet$, which in turn is interpreted as set equality between hereditarily finite hybrid sets.
Such equality is regulated by the standard \emph{extensionality axiom}, which has been proved to be equivalent, for hereditarily finite sets, to the following equational axioms \cite{Dovier00}:
\begin{align}
& \{x, x \plus E\} = \{x \plus E\} \tag{$Ab$} \label{Ab} \\
& \{x, y \plus E\} = \{y, x \plus E\} \tag{$C\ell$} \label{Cl}
\end{align}
Axiom \eqref{Ab} states that duplicates in a set term do not matter
(\emph{Absorption property}). Axiom \eqref{Cl} states that the order of
elements in a set term is irrelevant (\emph{Commutativity on the left}).
These two properties capture the intuitive idea that, for instance, the set terms $\{1,2\}$, $\{2,1\}$, and $\{1,2,1\}$ all denote the same set.

A \emph{valuation} $\sigma$ of a formula $\Phi$ is an assignment of
values from $D$ to the free variables of $\Phi$ respecting the sorts
of the variables. $\sigma$ can be extended to terms in a straightforward manner.
In the case of formulas, we write $\Phi[\sigma]$ to denote the
application of a valuation to a formula $\Phi$. $\sigma$ is a \emph{successful valuation} (or, simply, a \emph{solution}) if $\Phi[\sigma]$ is true in $\iS$.

\section{\label{expr}Expressing arrays in $\LQA$}

In this section we analyze the expressiveness of $\LQA$.
To begin with, many powerful set and relational operators can be defined as $\QA$-formulas. Table \ref{t:op} lists some of the most used operators definable in $\LQRI$. 
As can be seen $\Pi_\Set$ suffices to define all the Boolean operators of set theory.
The constraint $\Pfun$ mentioned in Section \ref{intro} is defined by means of RUQ as well as $\Ifun$ that captures the notion of injective function.
$\Ifun$ is expressible in $\LQA$ because it admits $\neq$-constraints in the $\mathit{QF}$-formulas, which is an improvement w.r.t. Bradley's results.

\begin{table}
\caption{\label{t:op}Some constraints definable in $\LQRI$}
\begin{minipage}{\columnwidth}
\begin{tabularx}{\columnwidth}{lXl}
\toprule
\textsc{Constraint} & \textsc{Definition} & \textsc{Interpretation}
\\\midrule
\multicolumn{3}{c}{\textsc{Integers}\footnote{Other order relations can be defined in a similar way.}} 
\\\midrule
$i \leq j$ &
$i < j \lor i = j$ &
standard
\\
$i > j$ &
$j < i$ & 
standard
\\\midrule
\multicolumn{3}{c}{\textsc{Sets}\footnote{$N,N_i$ are fresh variables.}} 
\\\midrule
$\Cap(E,F,G)$ & 
$\Cup(G,N_1,E) \land \Cup(G,N_2,F) \land  N_1 \disj N_2$ &
$G = E \cap F$
\\
$\Diff(E,F,G)$ &
$\Cup(E,G,E) \land \Cup(F,G,N_1) \land \Cup(E,N_1,N_1) \land F \disj G$ &
$G = E \setminus F$
\\
$E \subseteq F$ &
$\Cup(E,F,F)$ &
standard
\\\midrule
\multicolumn{3}{c}{\textsc{Binary Relations and Functions}} 
\\\midrule
$\Rel(R)$ &
$(\forall x \in R : \Pair(x))$ &
$R$ is a set of ordered pairs\footnote{In the following operators one can require $\Rel(R)$ whenever $R$ is an argument.}
\\
$\Pfun(R)$ &
$(\forall (x_1,y_1), (x_2,y_2) \in R : x_1 = x_2 \implies y_1 = y_2)$ &
$R$ is a function
\\
$\Ifun(R)$ &
$\Pfun(R) \land (\forall (x_1,y_1), (x_2,y_2) \in R : x_1 \neq x_2 \implies y_1 \neq y_2)$ &
$R$ is an injective function
\\
$\Dres(z,R,S)$ &
$\Cup(S,N,R) \land (\forall (x,y) \in S: x = z) \land (\forall (x,y) \in N: x \neq z)$ &
point-wise domain restriction
\\
$\Dres([k,m],R,S)$ &
$\Cup(S,N,R)$ &
domain restriction
\\
& ${}\land (\forall (x,y) \in S: k \leq x \leq m) \land (\forall (x,y) \in N: x < k \lor m < x)$ & 
\\[1mm]
$\Dares(z,R,S)$ &
$\Cup(S,N,R) \land (\forall (x,y) \in S: x \neq z) \land (\forall (x,y) \in N: x = z)$ &
{\small point-wise domain anti-restriction}
\\
$\Dares([k,m],R,S)$ &
$\Cup(S,N,R)$ &
domain anti-restriction 
\\
& ${}\land (\forall (x,y) \in S: x < k \lor m < x) \land (\forall (x,y) \in N: k \leq x \leq m)$ & 
\\\midrule
\multicolumn{3}{c}{\textsc{Arrays}}
\\\midrule
$\Arr(A,m)$ &
$\Pfun(A) \land \Size(A,m) \land (\forall (i,y) \in A: 1 \leq i \leq m)$ &
$A$ is an array of length $m$\footnote{In the following operators one can require $\Arr(A,m)$ whenever $A$ is an argument.}
\\
$\Get(A,i,y)$ &
$A = \{(i,y) \plus N\} \land (i,y) \notin N$ &
$A(i) = y$
\\
$\Upd(A,i,y,B)$ &
$A = \{(i,n) \plus N\} \land (i,n) \notin N \land B = \{(i,y) \plus N\}$ &
point-wise update\footnote{In Z, $B = A \oplus \{(x,y)\}$; in B, $B = A \lover \{(x,y)\}$; in Bradley, $B = A\{x \leftarrow y\}$.}
\\\midrule
\multicolumn{3}{c}{\textsc{Hashtables}}
\\\midrule
$\mathit{put}(H,k,v,T)$ &
$T = \{(k,v) \plus N\} \land \Dares(k,H,N) \land (k,v) \notin N$ &
\\
$\mathit{remove}(H,k,T)$ &
$\Dares(k,H,T)$ &
\\\bottomrule
\end{tabularx}
\end{minipage}
\end{table}

We provide two definitions for domain restriction ($\Dres$, cf.  equation \eqref{eq:dres}) in terms of primitive constraints of the language and RUQ, that apply to any binary relation $R$.
In both definitions the formulas assert the existence of a set, $N$, such that its union with $S$ yields $R$ and whose domains are disjoint---without defining the notion of domain.
The same is defined for domain anti-restriction (called $\Dares$ here but usually noted as $\ndres$), the complement of domain restriction w.r.t. $R$.
Basically, $\Dres$ and $\Dares$ are defined by using the fact that $R = (E \dres R) \cup (E \ndres R)$, where $E \dres R$ and $E \ndres R$ are obviously disjoint.
The first definition of $\Dres$ and $\Dares$ apply to an element of any sort; the second only to integer intervals.
$\LQA$ can also express range (anti-)restriction although they seem to be less useful.

\begin{remark}[Notation]
From now on we will use $A, B, C, D$ to name arrays.
\end{remark}

Given that arrays are sets of ordered pairs, all the operators defined in Table \ref{t:op} can be applied to them. For instance, we can perform the union of two arrays (although in general the result will not be an array), the subset relation can be used to check whether an array is a prefix of another one, and so on. Besides, although we present $\LQA$ as a language to work with arrays (i.e., a particular kind of functions), it can actually be used to write formulas involving binary relations, functions, integer intervals and set cardinality which goes far beyond arrays.

Although not shown in Table \ref{t:op}, it is worth to be mentioned that equality between arrays is just (extensional) set equality. That is, it is not necessary to define equality between arrays because they are encoded as sets of ordered pairs whose equality is standard (extensional) set equality. In turn, set equality is obtained by set unification \cite{Dovier2006} which is at the very base of all the previous works by the authors.

Concerning the definition of arrays in terms of set theory (i.e., $\Arr$ in Table \ref{t:op}), note that a more usual definition is the following.\footnote{This is an adaptation to arrays of the definition of sequences given in the Z specification language \cite{Spivey00}.}
\begin{equation}
\Arr(A,m) \iff \Pfun(A) \land \Dom(A) = [1,m]
\end{equation}
However, domain cannot be included in $\LQA$ without compromising its decidability \cite{DBLP:journals/tcs/CantoneL14,DBLP:journals/jar/CristiaR20}. 
Hence, we have to rest on the definition given in Table \ref{t:op} which uses $m$ to set $A$'s cardinality and a RUQ to state that $A$'s first components belong to $[1,m]$. Since $A$ is a function of cardinality $m$, its domain has the same cardinality and so, by the RUQ, the domain is equal to $[1,m]$. 
 
$\Get$ and $\Upd$ (read `update') represent the two main operators of any theory of arrays. $\Get(A,i,y)$ states that $y$ is the element stored at index $i$, whereas $\Upd$ modifies the array at index $i$. Note that our definitions are slightly more general than usual because they apply to functions in general.
Note that both $\Get$ and $\Upd$ fail if $i$ is not in $A$'s domain, although they do not fail if $A$ is not a function.
When used with arrays this will not be a problem.

Bradley (Section 7.3) shows how an assertion language for hashtables can be defined in terms of their theory.
In doing so they define three basic operators for hashtables that can be easily defined in $\LQA$---see the bottom of Table \ref{t:op}; the third operator is $\Get$.
Furthermore, we do not need to change anything in $\LQA$ nor in its solver to support hashtables.
In fact, in our theory a hashtable is just a function (set of ordered pairs) not constrained by the $\Arr$ predicate. Hence, all of our previous results apply easily to hashtables.

Clearly, more operators and specifications can be written in $\LQRI$. For instance, $\mathit{sorted}$ as defined in \eqref{sorted} can be written in $\LQRI$ as well as the loop invariant given in \eqref{inv_sorted} becomes the $\LQRI$ formula $\Dres([1,k],A,N) \land \mathit{sorted}(N)$.
Besides, $\LQRI$ admits specifications not definable in $\LINT$ nor in $\LRQ$ such as:
\begin{equation}
\Dres([1,k],A,A_l) \land \Dres([k+1,n],A,A_r) \land (\forall (i,y) \in A_l: y < 0) \land (\forall (i,y) \in A_r: y \geq 0)
\end{equation}
which partitions $A$ into a prefix of negative numbers and a suffix of non-negative numbers.

In particular, $\LQA$ admits all the specifications given by Bradley, as shown in Table \ref{t:bradley}. Bradley defines the \emph{array property} as the formulas of the form: $(\forall \vec{i}) (\varphi_I(\vec{i}) \rightarrow \varphi_V(\vec{i}))$ where $\vec{i}$ is a vector of index variables, and $\varphi_I(i)$ and $\varphi_V(i)$ are formulas of a particular kind.
$\varphi_I(i)$ can be a disjunction or conjunction of atoms of the form $i \leq j$ or $i = j$ where $i$ and $j$ are linear expressions not involving universally quantified array indexes. In turn, any $i \in \vec{i}$ in $\varphi_V(i)$ must be a read into an array.
Clearly, $\LQA$ can express any array property as defined by Bradley and more.
For instance, $\Ifun$ and $sorted$ (as defined in \eqref{sorted}) are not expressible in Bradley's language---Bradley can express sortedness but they need to refer to some array interval.
Note that indexes in Bradley's arrays start at 0 whereas in $\LQA$ they start at 1.
However, a more sensible difference is that in $\LQA$ the array length is explicit and finite whereas Bradley's arrays are infinite.
However, in Bradley's Section 7.2, some programs annotations include atoms of the form $\Card{a} = \Card{a_0}$, denoting the length of arrays, which seem not to be part of their language as given in Section 2.
Hence, $\LQA$ allows for overflow analysis over arrays, an important source of security vulnerabilities. Such an analysis amounts to just checking that every array read is within bounds. For example, in \Verb+sorted+ (Section \ref{intro}) we can simply check $1 \leq k \leq n \land 1 \leq k + 1 \leq n$ given that $\Arr(A,n)$ is part of the pre-condition.

\begin{table}
\caption{\label{t:bradley}Encoding of Bradley's examples in $\LQA$}
\begin{tabularx}{\columnwidth}{lX}
\toprule
\multicolumn{2}{c}{\textsc{Equality}} \\
\textsc{Bradley's} & $(\forall i)(a[i] = b[i])$ \\
$\LQA$ & $a = b$
\\\midrule
\multicolumn{2}{c}{\textsc{Bounded Equality}} \\
\textsc{Bradley's} &
  $(\forall i)(\ell \leq i \leq u \rightarrow a[i] = b[i])$ \\
$\LQA$ &
  $\Dres([\ell,u],a,N) \land \Dres([\ell,u],b,N)$
\\\midrule
\multicolumn{2}{c}{\textsc{Sorted}} \\
\textsc{Bradley's} &
   $\mathsf{sorted}(\ell,u,a) \defs (\forall i,j)(\ell \leq i \leq j \leq u \rightarrow a[i] \leq a[j])$
\\
$\LQA$ &
  $\Dres([\ell,u],a,N) \land (\forall (i_1,j_1),(i_2,j_2) \in N: i_1 \leq i_2 \implies j_1 \leq j_2)$
\\\midrule
\multicolumn{2}{c}{\textsc{Partitioned}} \\
\textsc{Bradley's} &
  $(\forall i,j)(\ell_1 \leq i \leq u_1 \leq \ell_2 \leq j \leq u_2 \rightarrow a[i] \leq a[j])$ \\
$\LQA$ &
  $\Dres([\ell_1,u_1],a,N_1) \land u_1 \leq \ell_2 \land \Dres([\ell_2,u_2],a,N_2) \land (\forall (i_1,j_1) \in N_1, (i_2,j_2) \in N_2: j_1 \leq j_2)$
\\\midrule
\multicolumn{2}{c}{\textsc{Array Property Formula} }\\
\textsc{Bradley's} &
  $(\exists\; \textsf{array } a)(\exists w,x,y,z,k,\ell,n \in \num)$ \\
& $\qquad (w < x < y < z \land 0 < k < \ell < n \land \ell - k > 1$ \\
& $\qquad{}\land \textsf{sorted}(0,n-1,a\{k \leftarrow w\}\{\ell \leftarrow x\}) \land \textsf{sorted}(0,n-1,a\{k \leftarrow y\}\{\ell \leftarrow z\})$ \\
$\LQA$ &
$w < x < y < z \land 0 < k < \ell \leq n \land \ell - k > 1$ \\
& ${}\land \Arr(a,n) \land \Upd(a,k,w,B) \land \Upd(B,\ell,x,B_1) \land \Upd(a,k,y,C) \land \Upd(C,\ell,z,C_1)$ \\
& ${}\land \mathit{sorted}(B_1) \land \mathit{sorted}(C_1)$
\\
\bottomrule
\end{tabularx}
\end{table}

Now let's look at what $\LQA$ \emph{cannot} express.
It cannot express a more general $\Upd$ operator updating an array in a set of indexes as it is available in B (i.e., $\lover$) and Z (i.e., $\oplus$). This definition would require to introduce either composition of functions or existential quantifiers, both of which compromise decidability \cite{DBLP:journals/tcs/CantoneL14,DBLP:journals/jar/CristiaR20,DBLP:journals/jar/CristiaR24}. The lack of a general $\Upd$ operator will render some specifications inexpressible in $\LQA$.
However, this would not be a severe impediment when dealing with loop invariants as loops usually update one element (or a fixed number of them) per iteration. The same limitation exists in Bradley's \emph{array property fragment} (APF).
Bradley explicitly rules out nested array reads, e.g., expressions such as $a[b[i]]$. In $\LQA$ this limitation is not explicit but is nonetheless enforced. Indeed, nested array reads can be naturally encoded, again, with composition. Actually the term $a[b[i]]$ is a point-wise instance of the composition between $a$ and $b$,  noted $a ; b$ and defined as the set $\{(x,z) \mid \exists y( (x,y) \in b \land (y,z) \in a)\}$. Since $\LQA$ does not admit composition nested reads are not expressible.
As a side note, see that \emph{secure} array composition is somewhat more complex as indexes must be within bounds, which in turn requires to know the length of arrays.

\begin{table}
\caption{\label{t:neg}Negation in $\LQA$}
\begin{minipage}{\columnwidth}
\begin{tabularx}{\columnwidth}{llll}
\toprule
\textsc{Constraint} & \textsc{Negation} & 
\textsc{Constraint} & \textsc{Negation}
\\\midrule
\multicolumn{4}{c}{\textsc{Integers}\footnote{Other negations can be defined in a similar way.}} 
\\\midrule
$i \leq j$ & $j < i$ &
$i = j$ & $i < j \lor j < i$ 
\\\midrule
\multicolumn{4}{c}{\textsc{Sets}\footnote{$n$ is a fresh variable.}} 
\\\midrule
$\Cup(E,F,G)$ & $n \in E \land n \notin G \lor n \in F \land n \notin G \lor n \in G \land n \notin E \land n \notin F$ &
$A \subseteq B$ & $n \in A \land n \notin B$
\\
$\Diff(E,F,G)$ & $n \in E \land n \notin F \land n \notin G \lor n \in G \land (n \notin E \lor n \in F)$ &
$\Size(A,k)$ & $\Size(A,n) \land n \neq k$ 
\\
$\Cap(E,F,G)$ & $n \in E \land n \in F \land n \notin G \lor n \in G \land (n \notin E \lor n \notin F)$ &
$E \disj F$ & $n \in E \land n \in F$
\\\bottomrule
\end{tabularx}
\end{minipage}
\end{table}

\begin{remark}[Negation]
As can be seen, what it is usually defined as operators in other approaches, in $\LQA$ becomes a predicate (in the form of constraints). 
For example, set union is usually defined as an operator taking two sets and returning another set but in $\LQA$ it is a predicate.
Then, for instance, if we want to state that $C$ is different from the union between $A$ and $B$ we have to negate $\Cup(A,B,C)$.
As in our previous works (e.g., \cite{Dovier00,DBLP:journals/tocl/CristiaR24}), negation is provided through $\LQA$ formulas encoding the negation of primitive constraints as shown in Table \ref{t:neg}.
In turn these formulas can be given a suitable name or symbol. For instance, the negation of $\Cup$ is called $\Ncup$ and the negation of $\subseteq$ is noted $\not\subseteq$.

On the other hand the negation of a RUQ of the form $(\forall x \in E:\phi)$ is given by the $\LQA$ formula: $n \in E \land \lnot\phi[x \mapsto n]$, where $n$ is a fresh variable and $\lnot\phi$ is the negation of $\phi$. As $\phi$ is a $\mathit{QF}$-formula its negation is easily computed.
For this reason, the negation of the relational and array constraints shown in Table \ref{t:op} can be easily obtained as $\LQA$ formulas.
\end{remark}

\section{\label{solver}$\SQA$, A CONSTRAINT SOLVER FOR $\LQA$}

\newcommand{\minsol}{\mathsf{check\_all\_minsol}}

$\SQA$, the solver for $\LQA$, is a combination of $\SATINT$ (read `sat-interval'), the solver for $\LINT$, and $\SATRQ$, the solver for $\LRQ$, plus some additions.
The overall organization of $\SQA$ is shown in Algorithm \ref{glob}.
Basically, $\SQA$ uses four routines: \textsf{gen\_size\_leq},
$\mathsf{STEP}$ (called from $\mathsf{step\_loop}$\footnote{As
$\mathsf{step\_loop}$ merely loops  calling $\mathsf{STEP}$, we will talk about the latter rather than the former. $\mathsf{STEP}$ is a key procedure in Algorithm \ref{glob}.}), \textsf{remove\_neq} and $\minsol$.
The key procedures involved in this work are $\mathsf{STEP}$ and $\minsol$.

\begin{algorithm}
\begin{minipage}{\textwidth}
\begin{algorithmic}[0]
 \State $\Phi \gets \textsf{gen\_size\_leq}(\Phi)$;
 \Repeat
   \State $\Phi' \gets \Phi$;
   \State $\Phi \gets \textsf{remove\_neq}(\mathsf{step\_loop}(\Phi))$
 \Until{$\Phi = \Phi'$; \hfill{\footnotesize[end of main loop]}}
 \If{$\Size \in \Phi$}
   \State\Return{$\minsol(\Phi)$}
 \Else
 \State\Return{$\Phi$}
 \EndIf
\vspace{2mm}
\Procedure{$\mathsf{step\_loop}$}{$\Phi$}
  \State \textbf{if} $\Phi = \Phi_1 \lor \Phi_2$ \textbf{then}
    \Return{$\mathsf{step\_loop}(\Phi_1) \sqcap \mathsf{step\_loop}(\Phi_2)$}
    \textbf{end if}
   \Repeat
     \State $\Phi' \gets \Phi$;
     \State $\Phi \gets \mathsf{STEP}(\Phi)$
     \hfill{\footnotesize[$\mathsf{STEP}$ is a key procedure]}
   \Until{$\Phi = \Phi'$}
 \State\Return{$\Phi$}
\EndProcedure
\end{algorithmic}
\hrule
\caption[fragile]{The solver $\SQA$. $\Phi$ is the input formula. $p \sqcap q$ means non-deterministically\footnote{The notion of non-deterministically executing $p$ or $q$ is that of CLP. That is, after $p$ is executed, in case the execution of $p$ fails, the system backtracks and executes $q$.} either $p$ or $q$.} \label{glob}
\end{minipage}
\end{algorithm}

As we will show in Section \ref{proofs}, when all the nondeterministic computations of $\SQA(\Phi)$ return $\false$, then we can conclude that $\Phi$ is unsatisfiable;
otherwise, when at least one of  them does not return $\false$, then we can conclude that $\Phi$ is satisfiable.

\textsf{gen\_size\_leq} simply adds integer constraints to the input formula $\Phi$ to force the second argument of $\Size$-constraints in $\Phi$ to be non-negative integers.
As can be seen, \textsf{step\_loop} considers the case when $\Phi$ is a disjunction by recursively and non-deterministically calling itself on each disjunct. Indeed, the semantics of the $\sqcap$ construct is the non-deterministic execution of both of its arguments.
$\mathsf{STEP}$ applies specialized rewriting procedures (Section \ref{rwrules}) to the current formula $\Phi$ and returns either $\false$ or the modified formula.
Each rewriting procedure applies a few nondeterministic rewrite rules which reduce the syntactic complexity of $\QA$-constraints of one kind.
\textsf{remove\_neq} deals with the elimination of $\neq$-constraints involving set variables, as is further described in Remark \ref{neqelim}.
$\minsol$ makes a final judgment when, after the main loop, $\Phi$ still contains $\Size$ constraints.

Broadly speaking, $\SQA$ works as follows.
The main loop replaces any defined constraint (i.e., those discussed in Section \ref{expr}) by its corresponding $\LQA$ formula and then it processes the remaining constraints---as explained in Section \ref{rwrules}.
The execution of $\mathsf{STEP}$ and \textsf{remove\_neq} is iterated until a fixpoint is reached, i.e., the formula is irreducible. 
These routines return $\false$ whenever (at least) one of the procedures in it rewrites $\Phi$ to $\false$. In this case, a fixpoint is immediately detected.
When the main loop terminates (Section \ref{termination}) the resulting $\Phi$ is either $\false$, in which case the input formula is unsatisfiable, or a formula in a particular form where, roughly (see Section \ref{irrcons}):
\begin{enumerate*}
\item Set arguments of $\Cup$, $\disj$ and $\Size$ constraints are variables;
\item There are no $\in$ constraints;
\item All the $=$ constraints involving set terms are of the form $var = term$; and
\item The domain of any RUQ is a variable.
\end{enumerate*}
If $\Phi$ still contains $\Size$ constraints, $\minsol$ is called to check whether or not a finite number of possible solutions, called \emph{minimum solutions}, satisfy $\Phi$.
If at least one minimum solution satisfies $\Phi$, then the input formula is surely satisfiable and a solution can be produced; if not, we claim that the input formula is unsatisfiable, basically, because any other model would include at least one minimum solution.
$\minsol$ is described in detail in Section \ref{after}.

\begin{remark}[$\SQA$ vs. $\SATINT$]\label{sqa_satint}
$\SQA$ is almost the same as $\SATINT$. The difference lays in the fact that $\LQA$ admits RUQ while $\LINT$ does not and so $\SQA$ must accommodate them. Therefore, \textsf{STEP} in $\SQA$ includes the rewrite rules dealing with RUQ (as defined in $\SATRQ$) and $\minsol$ processes them when performing the final judgment.
The similarity between $\SQA$ and $\SATINT$ is important because it simplifies the proof of correctness of $\SQA$.
\end{remark}

\subsection{\label{rwrules}Rewrite rules}
The core of Algorithm \ref{glob} is procedure \textsf{STEP} which in turn is composed of around 50 rewrite rules for set constraints grouped in seven rewrite procedures, one for each main set constraint. Most of the rules are inherited from $\SATINT$, a couple of them from $\SATRQ$ and one rule is specific of $\SQA$.
The rewrite rules used by $\SQA$ are defined as follows.

\begin{definition}[Rewrite rules]\label{d:rw_rules}
If $\pi$ is a symbol in $\Pi$ and $\phi$ is a $\QA$-constraint based on $\pi$, then a \emph{rewrite rule for $\pi$-constraints} is a rule of the form $\phi \lfun \Phi_1 \lor \dots \lor \Phi_m$, where $\Phi_i$, $1 \leq i \leq m$, are $\QA$-formulas.
If $p$ is a $\Sigma_{\QA}$-predicate and $\phi$ matches $p$, then $p$ is nondeterministically rewritten to one of the $\Phi_i$.
Variables appearing in the right-hand side but not in the left-hand side are assumed to be fresh variables, implicitly existentially quantified over each $\Phi_i$.
A \emph{rewriting procedure} for $\pi$-constraints consists of the collection of all the rewrite rules for $\pi$-constraints.
\end{definition}

Figure \ref{f:clpset} lists some representative rewrite rules inherited from $\SATINT$---the complete list is available in an online document \cite{calculusBR}.
Rule \eqref{eq:ext} is the main rule of set unification \cite{Dovier2006}. It states when two non-empty, non-variable sets are equal by nondeterministically and recursively computing four cases.
These cases implement the \eqref{Ab} and \eqref{Cl} axioms shown in Section \ref{semantics}. As an example, by applying rule \eqref{eq:ext} to $\{1\} = \{1,1\}$ we get:
$(1 = 1 \land \e = \{1\}) \lor (1 = 1 \land \{1\} = \{1\}) \lor (1 = 1 \land  \e = \{1,1\}) \lor (\e = \{1 \plus N\} \land  \{1 \plus N\} = \{1\})$, which turns out to be true (due to the second disjunct).

\begin{figure}
\hrule
\begin{align}
& \{x  \plus{} E\} = \{y \plus F\} \lfun \label{eq:ext}  \\
& \qquad  (x = y \land E = F)
   \lor (x = y \land \{x \plus E\} = F)
   \lor (x = y \land E = \{y \plus F\})
   \lor (E = \{y \plus N\} \land \{x \plus N\} = F) \notag \\[1mm]
& x \in E \lfun E = \{x \plus N\} \land x \notin N 
  \label{in:var} \\[1mm]
& x \in \{y \plus E\} \lfun x = y \lor x \in E \label{in:ext} \\[1mm]
& x \notin \{y \plus E\} \lfun x \neq y \land x \notin E
  \label{eq:in} \\[1mm]
& \text{If $G$ is a variable:} \label{un:ext1} \\
& \Cup(\{x \plus E\}, F, G) \lfun \notag \\
  & \qquad  \{x \plus E\} = \{x \plus N_1\}
      \land x \notin N_1 \land G = \{x \plus N\} \notag \\
  & \qquad \land (x \notin F \land \Cup(N_1,F,N)
  \lor F = \{x \plus N_2\}
       \land x \notin N_2 \land \Cup(N_1,N_2,N)) \notag \\[1mm]
& \Size(\emptyset,m) \lfun m = 0 \label{eq:size0}\\[1mm]
& \Size(E,0) \lfun E = \emptyset \label{eq:sizeempty} \\[2mm]
& \Size(\{x \plus E\},m) \lfun
    (x \notin E \land m = 1 + n \land \Size(E,n) \land 0 \leq n)
   \lor (E = \{x \plus N\} \land x \notin N \land \Size(E,m))
       \label{size:ext} \\[1mm]
& [k,m] = \{y \plus E\} \lfun \{y \plus E\} \subseteq [k,m] \land
   \Size(\{y \plus E\},m-k+1) & \label{e:ext} \\[1mm]
& \Cup(E,F,[k,m]) \lfun
 (m < k \land E = \e \land F = \e)
 \lor (k \leq m
       \land N \subseteq [k,m] \land \Size(N,m-k+1) \land \Cup(E,F,N))
         \label{un:3}
\end{align}
 \hrule
 \caption{\label{f:clpset}Some key rewrite rules inherited from $\SATINT$. $N,N_i,n$ are fresh variables.}
\end{figure}

Rule \eqref{un:ext1} is one of the main rules for $\Cup$-constraints.
It deals with constraints where the first argument is an extensional set and the last one a variable.
This rule uses set unification (i.e., rule \eqref{eq:ext}) and computes two cases: $x$ does not belong to $F$, and $x$ belongs to $F$ (in which case $F$ is of the form $\{x \plus N_2\}$ for some set $N_2$). In the latter, $x \notin N_2$ prevents Algorithm \ref{glob} from generating infinite terms denoting the same set. 

Rules \eqref{eq:size0}-\eqref{size:ext} are all but one of the rules dealing with $\Size$ constraints. Rule \eqref{size:ext} computes the size of any extensional set by counting the elements that belong to it while taking care of avoiding duplicates. For instance, the first nondeterministic choice for a formula such as $\Size(\{1,2,3,1,4\},m)$ will be:
\[
1 \notin \{2,3,1,4\}
  \land m = 1 + n \land \Size(\{2,3,1,4\},n) \land 0 \leq n
\]
which will eventually lead to a failure due to the presence of  $1 \notin \{2,3,1,4\}$ (see rule \eqref{eq:in}). This implies that $1$ will be counted in its second occurrence. Besides, the second choice becomes $\Size(\{2,3,1,4\},m)$ which is correct given that $\Card{\{1,2,3,1,4\}} = \Card{\{2,3,1,4\}}$.

Rule \eqref{e:ext} is based on the following known result:
\begin{equation}\label{eq:id}
k \leq m \implies (E = [k,m] \iff E \subseteq [k,m] \land \Card{E} = m - k + 1)
\end{equation}
Hence, the rule decides the satisfiability of $[k,m] = \{y \plus B\}$ by deciding the satisfiability of $\{y \plus B\} \subseteq [k,m] \land \Size(\{y \plus B\},m-k+1)$. Observe that \eqref{eq:id} is correctly applied as $k \leq m$ is implicit because $\{y \plus B\}$ is a non-empty set.

Finally, rule \eqref{un:3}, based again on \eqref{eq:id}, is crucial
as it permits to reconstruct an integer interval from the union of two sets. For instance, this rule covers constraints such as $\Cup(\{x \plus A\},B,[k,m])$, where all the arguments are variables.
Note that rule \eqref{un:3} eliminates the interval from the $\Cup$-constraint.

All the rules dealing with RUQ are shown in Figure \ref{f:ruq}, where the first three are inherited from $\SATRQ$ and the last one is specific of $\SQA$.
Rules \eqref{forall:iter1} and \eqref{forall:iter2} iterate over all the elements of the domain of the RUQ until it becomes the empty set or a variable.
In each iteration the unification between the quantified term and one element of the domain is attempted.
The unification is made explicit to cover the case when $c$ is an ordered pair.
If the unification succeeds (rule \eqref{forall:iter1}), the bound variables in $\phi$ are substituted by the result of the unification and a new iteration is fired.
Rule \eqref{forall:iter2} takes care of the case when the unification fails.
Intuitively, these rules transform a RUQ into a conjunction of ($\sInt$+$\sPair$+$\sUr$)-formulas.
For instance, $(\forall (x,y) \in \{1,(5,3)\}: \phi)$ is rewritten into $\phi[x \mapsto 5,y \mapsto 3]$ because $1$ is ignored by \eqref{forall:iter2} and rule \eqref{forall:iter1} succeeds when $(5,3)$ is processed.
In turn, rule \eqref{forall:int} is again based on \eqref{eq:id} thus turning a quantification over an interval into a quantification over an extensional set.
Note that the case $m < k$ covers the quantification over an empty domain.
As can be seen, \eqref{eq:id} is key to deal with integer intervals as sets.

\begin{figure}
\hrule
\begin{align}
& (\forall x \in \e: \phi) \lfun \true \label{forall:empty} \\[1mm]
& \text{If $\mathit{unify}(t,x)=\sigma$:} \label{forall:iter1} \\
& (\forall x \in \{t \plus E\}: \phi) \lfun
    \phi^\sigma \land (\forall x \in E: \phi) \notag \\[1mm]
& \text{If $\mathit{unify}(t,x)=\mathsf{failure}$:} \label{forall:iter2} \\
& (\forall x \in \{t \plus E\}: \phi) \lfun
    (\forall x \in E: \phi) \notag \\[1mm]
& (\forall x \in [k,m] : \phi) \lfun
  (k \leq m
     \land \Size(N,m-k+1)
     \land N \subseteq [k,m] \land (\forall x \in N : \phi))
  \lor
  m < k \label{forall:int}
\end{align}
\hrule
\caption{\label{f:ruq}Rewrite rules for RUQ. $N$ is a fresh variable; $x$ denotes a $CT$ term as in Definition \ref{formulas}.}
\end{figure}

In rules \eqref{forall:iter1} and \eqref{forall:iter2} if the domain becomes a variable the RUQ is not processed any more because there is no rule to apply in this case.
Constraints that are rewritten by no rule are called \emph{irreducible}. Irreducible constraints are part of the final answer of $\mathsf{STEP}$ as explained in the following section.

\begin{remark}[$\neq$-elimination]\label{neqelim}
As we have said, \textsf{remove\_neq} deals with the elimination of $\neq$-constraints involving set variables. It is a procedure present in one of the earliest works by the second author \cite{Dovier00}. In this version the procedure extends that of $\SATINT$ \cite[Figure 7]{DBLP:journals/tocl/CristiaR24} as follows, for any set variable $E$ and set term $t$ ($n$ is a fresh variable):
\begin{equation}
\text{If $E$ is an argument of $\Cup$ or $\Size$ or the domain of a RUQ: } 
\quad
E \neq t \lfun
  (n \in E \land n \notin t) \lor (n \in t \land n \notin E)
     \label{neq_elim:set}
\end{equation}
\end{remark}

\subsection{\label{irrcons}Irreducible constraints}
When no rewrite rule applies to the current formula $\Phi$ the main loop of $\SQA$ terminates returning $\Phi$ as its result.
If the returned formula is not $\false$ it can be seen, without loss of generality, as $\Phi_\Set \land \Phi_\Int$, where $\Phi_\Int$ contains all (and only) integer constraints and $\Phi_\Set$ contains all other constraints and RUQ occurring in $\Phi$.
The following definition precisely characterizes the form of atomic constraints in $\Phi_\Set$.

\begin{definition}[Irreducible formula]\label{def:solved}
Let $\Phi$ be a $\QA$-formula,
$E$ and $E_i$ variables of sort $\sSet$,
$X$ a variable and $t$ a term not of sort $\sInt$,
$x$ a term of any sort,
$c$ a variable or a constant integer number,
and $k$ and $m$ are terms of sort $\sInt$.
A $\QA$-constraint $\phi$ occurring in $\Phi$ is \emph{irreducible} if it has one of the following forms:
\begin{enumerate}[label=(\alph*)]
\item\label{i:eq} $X = t$, and neither $t$ nor $\Phi \setminus \{\phi\}$ contains $X$;
\item\label{i:neq} $X \neq t$, and $X$ does neither occur in $t$ nor as an argument of $(\Cup,\Size)$-constraints nor as the domain of a RUQ, in $\Phi$;
\item $x \notin E$, and $E$ does not occur in $x$;
\item $\Cup(E_1,E_2,E_3)$, where $E_1$ and $E_2$ are distinct variables;
\item\label{irr:disj} $E_1 \disj E_2$, where $E_1$ and $E_2$ are distinct variables;
\item\label{irr:forall} $\forall x \in E: \psi$, where $\psi$ is a $\mathit{QF}$-formula.
\item\label{irr:interval} $E \subseteq [k,m]$, where $k$ or $m$ are variables;
\item $\Size(E, c)$, $c \neq 0$;
\end{enumerate}
A $\QA$-formula $\Phi$ is irreducible if it is $\true$ or if all its
$\QA$-constraints are irreducible.
\end{definition}

$\Phi_\Set$, as returned by $\SQA$ once it finishes its main loop, is an irreducible formula.
This fact can be checked by inspecting the rewrite rules presented in \cite{calculusBR} and those given in Figure \ref{f:ruq}.
This inspection is straightforward as there are no rewrite rules dealing with irreducible constraints and all non-irreducible form constraints are dealt with by some rule.

$\Phi_\Set$ is the formula returned by $\SATINT$ except for the inclusion of RUQ (form \ref{irr:forall} in Definition \ref{def:solved}). 
It is important to observe that the atomic constraints occurring in $\Phi_\Set$
are indeed quite simple. In particular:
\begin{enumerate*}[label=(\roman*)]
\item All extensional set terms occurring in the input formula have been removed, except those occurring at right-hand sides of $=$ and $\neq$ constraints; and
\item All integer interval terms occurring in the input formula have been removed, except those occurring at the right-hand side of irreducible $\subseteq$-constraints.
\end{enumerate*}
Thus, all (possibly complex) equalities and inequalities between set terms have been solved.
Furthermore, all set arguments of $\Cup$, $\disj$ and $\Size$ constraints and the domains of RUQ are variables.

\begin{remark}
The irreducible form of Definition \ref{def:solved} is the same as the one defined for $\LINT$ plus form \ref{irr:forall} \cite{DBLP:journals/tocl/CristiaR24}.
\end{remark}

\begin{remark}\label{sub_int}
Concerning the irreducible constraints of form \ref{irr:interval} in Definition \ref{def:solved}, note that they can be written as RUQ:
\begin{equation}\label{eq:int_ruq}
E \subseteq [k,m] \iff (\forall x \in E: k \leq x \leq m)
\end{equation}
Moreover, no integer interval remains in an irreducible formula. This is important to simplify the proof of correctness of $\SQA$.
\end{remark}

\subsection{\label{after}Checking minimum solutions}
Due to the presence of $\Size$ and integer constraints, a non-$\false$ formula returned by the main loop is not always satisfiable, as is normally the case when these constraints are not present.

\begin{example}\label{ex:card_un}
Assuming all the arguments are variables, the following formula cannot be processed any further by $\mathsf{STEP}$ but is unsatisfiable:
\begin{equation*}
\Cup(E,F,G) \land \Size(E,m_e) \land \Size(F,m_f) \land \Size(G,m_g)
\land m_e + m_f < m_g
\end{equation*}
as it states that $\Card{E} + \Card{F} < \Card{E \cup F}$.
\end{example}

Therefore, once the main loop terminates, Algorithm \ref{glob} calls $\minsol$, which is described in Algorithm \ref{a:minsol}.
As can be seen, the first step divides $\Phi$ into two subformulas characterized as follows:
\begin{enumerate*}
\item $\Gamma_1$, a conjunction of all the integer constraints and
all the $\Cup$, $\disj$ and $\Size$ constraints; and
\item $\Gamma_2$, the rest of $\Phi$.
\end{enumerate*}

\begin{algorithm}
\begin{algorithmic}[0]
\Procedure{$\minsol$}{$\Phi$}
 \State \textbf{let} $\Phi$ 
        \textbf{be} $\Gamma_1 \land \Gamma_2$;
 \If{$\STEPCARD(\Gamma_1,Min)$}
 \If{$\forall \in \Gamma_2$}
 \For{$m_1 = c_1, \dots, m_k=c_k$ solution of $0 \leq m_1 \leq Min \land \dots \land 0 \leq m_k \leq Min \land Min = \sum_{i=1}^k m_i$}
    \If{$\mathsf{step\_loop}(\Phi \land m_1 = c_1 \land \dots \land m_k=c_k) \neq \false$}
       \State\Return{$\Phi$}
    \EndIf
    \EndFor
    \State\Return{$\false$}
    \Else \State\Return{$\Phi$}
    \EndIf
 \Else
    \State\Return{$\false$}
 \EndIf
\EndProcedure
\end{algorithmic}
\caption{\label{a:minsol}The $\minsol$ procedure. $\Phi$ is the formula produced by the main loop of Algorithm \ref{glob}.}
\end{algorithm}

Then, procedure $\STEPCARD$ is called with $\Gamma_1$ as input and, if the call succeeds, $Min$ represents its output.
$\STEPCARD$ implements a decision procedure for formulas such as $\Gamma_1$, i.e., formulas including LIA, $\Size$, $\Cup$ and $\disj$ constraints \cite{DBLP:journals/tplp/CristiaR23}.
LIA problems are solved by means of the SWI-Prolog CLP(Q) library \cite{holzbaur1995ofai} which provides the following predicate:
\[
\mathsf{bb\_inf}(\mathit{Vars,Expr,Min,Vert})
\]
which finds a vertex ($\mathit{Vert}$) of the minimum ($\mathit{Min}$) of the expression $\mathit{Expr}$ subjected to the integer constraints present in the constraint store and assuming all the variables in $\mathit{Vars}$ take integer values.
$\STEPCARD$ calls $\mathsf{bb\_inf}$ as follows:
\[
\mathsf{bb\_inf}(\mathit{intVars_{\Gamma_1},\sum_{i=1}^k m_i,Min,\_})
\]
where $intVars_{\Gamma_1}$ are all the integer variables present in $\Gamma_1$;
each $m_i$ is the second argument of a $\Size$ constraint present in $\Gamma_1$;
and $Min$ is a new variable.
That is, $\mathsf{bb\_inf}$ is called to minimize the sum of all the cardinalities present in $\Gamma_1$. 
However, notice that this minimization succeeds only if the integer constraints present in $\Gamma_1$ are satisfiable; if not, $\mathsf{bb\_inf}$ simply fails, making $\STEPCARD$ to fail as well.
If this call to $\mathsf{bb\_inf}$ succeeds then $Min$ is bound to an integer number.

If $\STEPCARD(\Gamma_1,Min)$ succeeds and there is a RUQ in $\Gamma_2$, $\minsol$ iterates over all the solutions of the integer formula:
\begin{equation}\label{e:clpfd}
0 \leq m_1 \leq Min \land \dots \land 0 \leq m_k \leq Min \land Min = \sum_{i=1}^k m_i
\end{equation}
A solution to the above formula is called \emph{minimum solution}.
This is because the cardinalities present in $\Gamma_1$ can only assume values less than or equal to $Min$, which in turn is the minimum value of the sum of all cardinalities.
When \eqref{e:clpfd} is solved $Min$ is an integer number, not a variable.
Furthermore, in the minimum solution $m_1 = c_1, \dots, m_k=c_k$ each $c_i$ is an integer number.

\begin{remark}\label{minsol}
It is easy to see that there is a finite number of minimum solutions. \end{remark}

\begin{remark}
In $\SATINT$ the condition to execute the \textbf{for} loop is $E \subseteq [m,k] \in \Gamma_2$.
That is, $\SQA$ generalizes that condition to the presence of any RUQ.
\end{remark}

Therefore, $\minsol$ checks whether or not any minimum solution is a solution of $\Phi$.
As soon as a minimum solution is a solution of $\Phi$, $\minsol$ terminates returning $\Phi$.
If no minimum solution is a solution of $\Phi$, $\minsol$ returns $\false$ meaning that $\Phi$ is unsatisfiable.
In the later case we claim that the input formula is unsatisfiable, basically, because any other model would include a minimum solution.
This is proved in Section \ref{proofs}.

Observe that if $m_1 = c_1, \dots, m_k=c_k$ is a minimum solution then all the cardinalities present in $\Phi$ are bound to integer numbers.
Then, all $\Size$ constraints in $\Phi$ become of the form $\Size(E,c)$ with $E$ a variable and $c$ an integer number.
In this case the following rewrite rule is activated:
\begin{equation}
\text{If $k$ is an integer number: } \qquad
\Size(E,k) \lfun
  E = \{n_1,\dots,n_k\}
  \land ad(n_1,\dots,n_k) \label{size:const3}
\end{equation}
where $n_1,\dots,n_k$ are fresh variables and $ad(n_1,\dots,n_k)$ is a shorthand for $\bigwedge_{i=1}^{k-1} \bigwedge_{j=i+1}^k n_i \neq n_j$ (i.e., all $n_i$ are different from each other).
Hence, $\minsol$ calls $\mathsf{step\_loop}$, instead of $\SQA$, because there is no need to recursively call $\minsol$ again as there will be no $\Size$ constraints after $\mathsf{step\_loop}$ applies rule \eqref{size:const3}.

\section{\label{proofs}$\SQA$ IS A DECISION PROCEDURE}
In this section we analyze the correcteness of $\SQA$ by considering its soundness, completeness and termination. This analysis is strongly based on previous results by the authors as is duly referenced whenever pertinent.

\subsection{Soundness and completeness}
The following theorem ensures that each rewrite rule used by $\SQA$
preserves the set of solutions of the input formula.

\begin{theorem}[Equisatisfiability]\label{equisatisfiable}
Let $\phi$ be a constraint based on symbol $\pi \in \Pi \setminus \PiInt$ or a RUQ, and $\phi \lfun \Phi_1 \lor \dots \lor \Phi_n$ a rewrite rule for $\pi$.
Then, each solution $\sigma$ of $\phi$ is a solution of $\Phi_1 \lor \dots \lor \Phi_n$, and vice versa, i.e., $\iS \models \phi[\sigma] \iff \iS \models (\Phi_1 \lor \dots \lor \Phi_n)[\sigma]$.
A solution of $\phi$ is expanded to the variables occurring in $\Phi_i$ but not in $\phi$, so as to account for the possible fresh variables introduced into $\Phi_i$.
\end{theorem}

\begin{proof}
The proof is based on showing that, for every rewrite rule, the set of solutions at the left-hand side of a rewrite rule is the same as the one at the right-hand side. Except for rule \eqref{forall:int}, the proofs of equisatisfiability for all the other rules have been done in earlier works by the authors \cite{Dovier00,DBLP:journals/tplp/CristiaR23,DBLP:journals/tocl/CristiaR24,DBLP:journals/jar/CristiaR24}.

The proof of equisatisfiability for rule \eqref{forall:int} is rather simple as it is based on \eqref{eq:id}. Indeed, if $k \leq m$ the interval $[k,m]$ is replaced by $N$ which is included in the interval and has the same cardinality, then $N$ is equal to $[k,m]$. On the other hand, if $m < k$ then $[k,m] = \emptyset$ in which case the RUQ becomes $(\forall x \in \emptyset: \phi)$ which is trivially true (rule \eqref{forall:empty}).
\end{proof}

The following is a consequence of the above theorem.

\begin{corollary}\label{coro}
Let $\Phi$ be the input formula to $\SQA$ and let $\Phi_1, \dots, \Phi_n$ be the formulas obtained at the end of the main loop of Algorithm \ref{glob}. Then, each solution $\sigma$ of $\Phi$ is a solution of some $\Phi_i$, and vice versa, i.e., $\iS \models \Phi[\sigma] \iff \iS \models \Phi_i[\sigma]$, for some $i$.
\end{corollary}

\begin{proof}
Putting RUQ aside, we have proved elsewhere that by the end of the main loop of Algorithm \ref{glob} the above holds \cite[Theorem 2]{DBLP:journals/tocl/CristiaR24}.
When RUQ are considered, Algorithm \ref{glob} adds the rewrite rules of Figure \ref{f:ruq} which, by Theorem \ref{equisatisfiable}, preserve solutions. Hence, the corollary is proved.
\end{proof}

From now on we will analyze properties of the formulas returned by the main loop of Algorithm \ref{glob}---i.e., each $\Phi_i$ mentioned in the above corollary. Recall that each one of these formulas is a conjunction of atomic predicates.

As explained in Section \ref{after}, $\STEPCARD$ is needed to decide the satisfiability of formulas including $\Size$ and integer constraints.
When $\STEPCARD$ is called from inside of $\minsol$, the $\Phi_i$ passed in to it is either $\false$ or a conjunction of atomic predicates of the form $\Phi_\Set \land \Phi_\Int$, where $\Phi_\Int$ contains all (and only) integer constraints and $\Phi_\Set$ contains all other constraints and RUQ occurring in $\Phi_i$.
Moreover, $\Phi_\Set$ is a conjunction of irreducible constraints enjoying the following property. 

\begin{lemma}\label{l:set-ruq}
Let $\Phi_i$ be a conjunction of irreducible constraints of forms \ref{i:eq}-\ref{irr:forall}---i.e., $\Phi_i$ is free of $\Size$ constraints and integer intervals due to Remark \ref{sub_int}. Then, $\Phi_i$ is satisfiable.
\end{lemma}

\begin{proof}
Set $\Phi_i \defs \Phi_{ae} \land \Phi_{f}$, where
$\Phi_{ae}$ is a conjunction of irreducible constraints of forms \ref{i:eq}-\ref{irr:disj} and
$\Phi_{f}$ is a conjunction of irreducible RUQ of form \ref{irr:forall}.
The earliest work by the authors \cite[Definition 9.1 and Theorem 9.4]{Dovier00} proves that formulas such as $\Phi_{ae}$ are satisfiable by substituting all set variables, except those in form \ref{i:eq}, by the empty set. On the other hand, observe that $\Phi_{f}$ is also satisfiable by substituting domains by the empty set.
Then, $\Phi_i$ is satisfiable.
\end{proof}

The following is a trivial corollary following from the previous lemma and the rewrite rules applied by $\SQA$.

\begin{corollary}\label{coro2}
Let $\Phi_i$ be a formula satisfying the conditions of Lemma \ref{l:set-ruq}, then $\mathsf{step\_loop}(\Phi_i)$ is either $\false$ or a formula satisfying the conditions of Lemma \ref{l:set-ruq}.
\end{corollary}

\begin{proof}
Given that  $\Phi_i$ is a conjunction of irreducible constraints of forms \ref{i:eq}-\ref{irr:forall}, $\textsf{STEP}(\Phi_i)$ cannot generate a $\Size$ constraint or an integer interval simply because $\Size$ constraints and integer intervals are generated only when other $\Size$ constraints and integer intervals are processed.
\end{proof}

If a formula $\Phi_i$ returned after the main loop is $\false$, the final answer of Algorithm \ref{glob} is $\false$ as the \textbf{else} branch of the conditional sentence is executed.
Hence, in order to prove the correctness of $\SQA$ we need to prove that the  \textbf{then} branch is correct. 

\begin{lemma}\label{l:main}
Let $\Phi_i$ be a formula returned after the main loop of Algorithm \ref{glob} different from $\false$. 
Then, $\Phi_i$ is satisfiable iff the formula returned by $\minsol(\Phi_i)$ is satisfiable.
\end{lemma}

\begin{proof}
By looking at $\minsol$ in Algorithm \ref{a:minsol} we know that the first step is calling $\STEPCARD(\Gamma_1,Min)$ where $\Gamma_1$ is an input and $Min$ is an output.
As we said in Section \ref{after},  $\STEPCARD$ implements a decision procedure for formulas like $\Gamma_1$.
Then, if $\STEPCARD(\Gamma_1,Min)$ returns $\false$ then $\Gamma_1$ is unsatisfiable and so $\Phi_i$.
Now we assume $\Gamma_1$ is satisfiable.
Since $\Gamma_1$ is a conjunction of LIA, $\Size$, $\Cup$ and $\disj$ constraints, we can think that it is of the form:
\begin{equation}\label{eq:gama1}
\Size(E_1,k_1) \land \dots \land \Size(E_a,k_a) \land \Gamma_0
\end{equation}
where $\Gamma_0$ is free of $\Size$ constraints.
In this case, when $\STEPCARD(\Gamma_1,Min)$ returns, we know that $Min$ is the minimum of $\sum_{i=1}^a k_i$ subjected to the integer constraints in $\Gamma_1$---as explained in Section \ref{after}.

Next, $\minsol$ either returns $\Phi_i$ or iterates over all the minimum solutions given by the LIA problem \eqref{e:clpfd}.
In each iteration we have the following call:
\begin{equation}\label{eq:minsol}
\mathsf{step\_loop}(\Phi_i \land k_1 = c_1 \land \dots \land k_a=c_a)
\end{equation}
where each $c_i$ is an integer number---as explained in Section \ref{after}.
Clearly, if $\Phi_i \land k_1 = c_1 \land \dots \land k_a=c_a$ is satisfiable so $\Phi_i$ simply because $\SQA$ has found a solution for it.
So we have proved that $\minsol$ does not return $\false$ when $\Phi_i$ is satisfiable.

However, the important part of this proof is when \eqref{eq:minsol} returns $\false$ for every minimum solution because in this case we argue that $\Phi_i$ is unsatisfiable in spite that it is evaluated on just a finite number of solutions of $\Gamma_1$---i.e., the minimum solutions.

In order to do that, we consider any given minimum solution given as $k_1 = c_1 \land \dots \land k_a=c_a$ and let us set $\Phi_i = \Gamma_1 \land \Gamma_2 \land \Gamma_3$ where:
\begin{enumerate}[label=(\roman*)]
\item\label{i:phi1} $\Gamma_1 \defs \Size(E_1,k_1) \land \dots \Size(E_a,k_a) \land \Gamma_0$, as in \eqref{eq:gama1}
\item\label{i:phi2} $\Gamma_2$ is a conjunction of all the $\notin$, $=$ and $\neq$ constraints not involving integer terms
\item\label{i:phi3} $\Gamma_3$ is a conjunction of RUQ (due to Remark \ref{sub_int})
\end{enumerate}
Notice that this is aligned with the irreducible form given in Defintion \ref{def:solved}.

Besides, as soon as $\mathsf{step\_loop}$ executes \eqref{size:const3}, the input formula in \eqref{eq:minsol} becomes:
\[
\Psi \land \Gamma_0 \land \Gamma_2 \land \Gamma_3
\]
where:
\[
\Psi \defs E_1 = \{n_1^1,\dots,n_{c_1}^1\} \land ad(n_1^1,\dots,n_{c_1}^1) \land \dots \land E_a = \{n_1^a,\dots,n_{c_a}^a\} \land ad(n_1^a,\dots,n_{c_a}^a)
\]

If \eqref{eq:minsol} returns $\false$ it is because $\Gamma_3$ is not satisfied by the valuation defined in $\Psi$.
In order to prove that, we will see that $\Gamma_2$ is satisfied by the valuation defined in $\Psi$.
Due to \ref{i:phi2} the constraints in $\Gamma_2$ are of the following forms:
\begin{itemize}
\item $x \notin E$, where $E$ is a variable. Then if $E$ is one $E_i$ in $\Psi$ the systematic application of rule \eqref{eq:in} yields a \emph{finite }conjunction of $\neq$-constraints of the form $x \neq n$ where $n$ is a fresh variable.
Now, since all the sorts are mapped to infinite domains, no finite conjunction of $\neq$-constraints can make the formula unsatisfiable.
\item $X \neq t$, where $X$ is a variable not occurring in $\Size$-constraints (item \ref{i:neq} in Definition \ref{def:solved}). So no substitution can occur given that all the $E_i$ in $\Psi$ were arguments of $\Size$-constraints.
\item $X = t$, where $X$ is a variable not occurring elsewhere in the  formula (item \ref{i:eq} in Definition \ref{def:solved}). So no substitution can occur.
\end{itemize}
As a consequence, $\Gamma_2$ is satisfied by the valuation defined in $\Psi$ and so if \eqref{eq:minsol} returns $\false$ it is because $\Gamma_3$ is not satisfied by the valuation defined in $\Psi$.

Now, let's see why if \eqref{eq:minsol} returns $\false$ $\Phi_i$ is unsatisfiable.
Let $\sigma$ be valuation of $\Gamma_1 \land \Gamma_2 \land \Gamma_3$.
If $\sigma$ coincides with the valuation defined in $\Psi$ for all $E_i$ then $\Gamma_1 \land \Gamma_2 \land \Gamma_3$ will not be satisfied---because we are assuming \eqref{eq:minsol} returns $\false$ for all minimum solutions.
Hence, we have to consider that there is at least one $E_{i_0}$ ($i_0 \in [1,a]$) in $\sigma$ bound to a set with at least one more element, i.e., $E_{i_0} = \{n_1^{i_0},\dots,n_{c_{i_0}}^{i_0},n_{c_{i_0}+1}^{i_0}\}$ under the condition given by $ad(n_1^{i_0},\dots,n_{c_{i_0}}^{i_0},n_{c_{i_0}+1}^{i_0})$.
Observe that $E_{i_0}$ cannot be smaller because $c_{i_0}$ is the minimum cardinality for $E_{i_0}$ as given by the considered minimum solution that in turn satisfies $\Gamma_1$.
Hence, if we consider a smaller $c_{i_0}$ then $\Gamma_1$ will not be satisfied, so we are constrained to consider a larger cardinality for $E_{i_0}$.
We may consider reducing $c_{i_0}$ while increasing some other constant (i.e., some other $c_i$) in this minimum solution but that would be another minimum solution.
We cannot consider a \emph{different} set for $E_{i_0}$ because all $n_j^i$ are (fresh) variables.

We know that $\Gamma_3$ is not satisfied by the valuation defined in $\Psi$. This means that there is one or more RUQ in $\Gamma_3$ whose domain are some $E_j$ (with $j \in J \subseteq [1,a]$). 
Then, these $E_j$ have been substituted by the corresponding sets defined in $\Psi$.
If $\sigma$ assigns to all these $E_j$ sets of the same cardinality as those given in $\Psi$ then $\Gamma_3$ will not be satisfied by $\sigma$ (because the same substitutions will take place in $\Gamma_3$).
Therefore, we have to consider that at least one of the $E_j$ is $E_{i_0}$. For now we will consider that there is just one RUQ in $\Gamma_3$ whose domain is $E_{i_0}$.

Note that all the RUQ in $\Gamma_3$ are such that their quantifier-free formulas do not contain free set variables. This means that these quantifier-free formulas remain the same after the substitutions induced by $\Psi$ or $\sigma$.
Now, at the left, we apply the substitution induced by $\Psi$ (i.e., $E_i = \{n_1^i,\dots,n_{c_i}^i\}$); and, at the right, the substitution induced by $\sigma$ (i.e., $E_i = \{n_1^i,\dots,n_{c_i}^i,n_{c_i+1}^i\}$):
\begin{align*}
& (\forall x \in E_i : \phi)[E_i \mapsto \{n_1^i,\dots,n_{c_i}^i,n_{c_i}^i\}]
& (\forall x \in E_i : \phi)[E_i \mapsto \{n_1^i,\dots,n_{c_i}^i,n_{c_i+1}^i\}] \why{property of RUQ} \\
& \equiv
\bigwedge_{j=1}^{c_i} \phi[x \mapsto n_j^i] 
& \equiv
\bigwedge_{j=1}^{c_i+1} \phi[x \mapsto n_j^i] \hspace{2.8cm}
  \why{property of $\land$} \\
& 
& \equiv
\phi[x \mapsto n_{c_i+1}^i] \land \bigwedge_{j=1}^{c_i} \phi[x \mapsto n_j^i] \hspace{.7cm}
\end{align*}
Recall that we have assumed that the substitution induced by $\Psi$ does not satisfy $\Gamma_3$.
On the other hand, see that the final formula obtained in the substitution induced by $\sigma$ contains $\bigwedge_{j=1}^{c_i} \phi[x \mapsto n_j^i]$ that is exactly the result of the substitution induced by $\Psi$---which in turn is possible because $\phi$ remains the same after each substitution, as we analyzed above.
Hence, $\Gamma_3$ cannot be satisfied by $\sigma$.

The same argument can be generalized to any number of $E_i$ with any number of elements in each of them and any number of RUQ whose domain is some $E_i$, because all of them contribute to a conjunction where a subformula is unsatisfiable.
Then, if every minimum solution given by the LIA problem \eqref{e:clpfd} is not a solution of $\Phi_i$, $\Phi_i$ is unsatisfiable.

Therefore, the lemma is proved.
\end{proof}

Next we prove that after termination $\SQA$ is sound and complete.

\begin{theorem}[Soundness and completeness]\label{sound&complete}
Let $\Phi$ be a $\QA$-formula and $\Phi_1, \dots,\Phi_n$ be the
collection of formulas returned by $\SQA(\Phi)$.
Then:
\begin{enumerate}
\item\label{i:sat} $\Phi$ is satisfiable iff at least one of $\Phi_1, \dots,\Phi_n$ is satisfiable. 
\item\label{i:sol} Every solution of $\Phi_1, \dots,\Phi_n$ is a solution of $\Phi$.
\end{enumerate}
\end{theorem}

\begin{proof}
Corollary \ref{coro} and Lemma \ref{l:main} prove \eqref{i:sat}.
For the proof of \eqref{i:sol} let $\Phi'_i$ be the formula at the end of the main loop corresponding to $\Phi_i$. By Corollary \ref{coro} we know that the set of solutions of $\Phi'_i$ is a subset of the solutions of $\Phi$. Then, $\minsol(\Phi')$ further restricts the set of solutions of $\Phi'_i$ by considering only the minimum solutions induced by $\Gamma_1$. Now we can apply Theorem \ref{equisatisfiable} and Corollary \ref{coro2}.
\end{proof}

\subsection{\label{termination}Termination}

If the formula passed in to Algorithm \ref{glob} is free of RUQ, then it is a $\LINT$ formula (Remark \ref{lint-lqa}). In this case, $\SQA$ behaves as $\SATINT$ as discussed in Remark \ref{sqa_satint}. Hence, termination is trivial as $\SATINT$ has been proved to be a decision procedure for $\LINT$ formulas.
Similarly, if the formula passed in to Algorithm \ref{glob} is a conjunction of $\QA$-RUQ, then $\SQA$ behaves as $\SATRQ$. $\SATRQ$ is guaranteed to terminate if the quantifier-free formulas inside RUQ are free of $\in$ constraints \cite[Theorem 3]{DBLP:journals/jar/CristiaR24}, which is the case of $\QA$-RUQ.
Hence, it remains to be proved that $\SQA$ terminates on formulas combining $\LINT$ formulas with $\QA$-RUQ.

\begin{lemma}\label{l:loopstepterm}
Let $\Phi$ be a conjunction of constraints and RUQ belonging to $\LQA$. Then, there is an implementation of \textsf{loop\_step} such that $\textsf{loop\_step}(\Phi)$ terminates.
\end{lemma}

\begin{proof}
The proof is based on considering that \textsf{loop\_step} processes all constraints first and RUQ the last. So let $\Phi \defs \Phi_c \land \Phi_r$ where $\Phi_c$ is a conjunction of constraints and $\Phi_r$ a conjunction of RUQ. When \textsf{loop\_step} terminates processing $\Phi_c$ it returns an irreducible formula containing forms \ref{i:eq}-\ref{irr:disj} (Definition \ref{def:solved}).

Next, \textsf{loop\_step} considers the irreducible equalities $X = t$ (form \ref{i:eq}). Say one of these equalities is the following:
\begin{equation}\label{eq:eq}
E = \{e \plus N\}
\end{equation}
These are the only constraints to be considered at this point because the remaining ones have no interaction with RUQ. Actually, this step has some influence on RUQ if at least one $X$ is a variable appearing in some RUQ. Then, say we have the following RUQ:
\begin{equation}
(\forall x \in E: \phi)
\end{equation}
Hence, \textsf{loop\_step} substitutes $X$ by $t$ in all RUQ. In this case we get the following:
\begin{equation}\label{eq:ruq1}
(\forall x \in \{e \plus N\}: \phi)
\end{equation}
 
Now \textsf{loop\_step} processes $\Phi_r$ which terminates---due to $\SATRQ$ being a decision procedure for RUQ, as discussed above. Furthermore, when this process terminates all RUQ are irreducible. So \eqref{eq:ruq1} becomes:
\begin{equation}\label{eq:ruq2}
(\forall x \in N: \phi)
\end{equation}
However, when \textsf{loop\_step} processes $\Phi_r$, zero or more $\mathit{QF}$ constraints are generated.
These constraints are processed again (as $\Phi_c$ above) thus yielding a new set of constraints, $C$.
As all the RUQ are irreducible the only chance to start another iteration of the solving algorithm is that some constraint in $C$ is of the form $F = \{\dots\}$, with $F$ the domain of some RUQ, which will fire a new substitution and a new round of RUQ solving.
Hence, we will analyze $\mathit{QF}$ constraints whose processing could generate constraints of the form $F = \{\dots\}$.
\begin{enumerate}
\item\label{i:seteq} $z = y$, where $z$ and $y$ are extensional set terms or set variables.
To be more illustrative $z = y$ can be of the form $\{e_1 \plus F_1\}  = \{e_2 \plus F_2\}$ where $F_i$ can be $E$, $N$ or some other variable.
In any case, when such an equality is solved by applying rule \eqref{eq:ext}, an equality such as \eqref{eq:eq} can be generated where, however, $E$ and $N$ cannot appear inside the set term at the right-hand side (due to form \ref{i:eq} given in Definition \ref{def:solved}). Hence, eventually Algorithm \ref{glob} halts.
\item $z \neq y$, where $z$ and $y$ are set terms.
That is, $z \neq y$ is of the form $\{e_1 \plus F_1\} \neq \{e_2 \plus F_2\}$. In this case, $\in$ and $\notin$ constraints are generated. $\notin$ constraints cannot affect RUQ domains. $\in$ constraints can become equalities such as \eqref{eq:eq} and so case \eqref{i:seteq} applies.
\item $z < y$, which has no influence on the domains of RUQ.
\end{enumerate}

\end{proof}

We close this section by proving the main termination theorem.
\begin{theorem}[Termination]
$\SQA$ can be implemented in such a way that it terminates for all formulas in $\LQA$.
\end{theorem}

\begin{proof}
The implementation of $\SQA$ is Algorithm \ref{glob} where \textsf{step\_loop} is implemented as in Lemma \ref{l:loopstepterm}.

If the formula to be solved is a disjunction then $\SQA$ solves one term at a time. Hence, we need to prove that $\SQA$ terminates for any conjunction of atoms and RUQ belonging to $\LQA$.

To that end we first need to prove that the main loop of Algorithm \ref{glob} terminates for such a formula.
Note that if the rewrite rule implemented in \textsf{remove\_neq} is not fired, then once \textsf{step\_loop} terminates, the main loop also terminates.
On the other hand, rule \eqref{neq_elim:set} is fired only when the formula is irreducible except for the presence of inequalities of the form $X \neq t$ where $X$ is an argument of ($\Cup$, $\Size$)-constraints or the domain of a RUQ. Once this rule is applied the $\neq$ constraint that fired it is eliminated from the formula making it impossible for the rule to be fired again.
Moreover, if rule \eqref{neq_elim:set} is fired it adds $\in$ and $\notin$ constraints to the formula which are known to not compromise termination of \textsf{step\_loop} \by{Lemma \ref{l:loopstepterm}}.
This proves that the main loop terminates.

Now we need to prove that the \textbf{if-then-else} statement following the main loop terminates as well.
The \textbf{else} branch trivially terminates.
The \textbf{then} branch terminates because $\minsol(\Phi)$ terminates because $\STEPCARD(\Gamma_1,Min)$ terminates as $\STEPCARD$ has been proved to be a decision procedure---as explained in Section \ref{after}---and the \textbf{for} loop also terminates because it iterates a finite number of times (see Remark \ref{minsol}) and in each iteration $\mathsf{step\_loop}$ is executed which in turn terminates (due to Lemma \ref{l:loopstepterm}).
\end{proof}

\section{\label{setlog}IMPLEMENTATION OF $\SQA$ IN \setlog}
\setlog is a publicly available tool \cite{setlog} implementing an advanced form of logic set programming \cite{DBLP:books/daglib/0067831,DBLP:series/mcs/CantoneOP01}.
The tool is a CLP language and satisfiability solver implemented in SWI-Prolog.
It offers a set-based language for the specification of state machines, a typechecker, a verification condition generator and a test case generator.
\setlog implements decision procedures for several fragments of set theory and set relation algebra and uses the CLP(Q) Prolog library to solve LIA problems.

In order to implement the ideas put forward in this paper we slightly modify the \setlog solver by adding rule \eqref{forall:int} and the condition to execute the \textbf{for} loop in Algorithm \ref{a:minsol} which now takes care of the presence of RUQ in the irreducible formula delivered by the main loop of Algorithm \ref{glob}.
Moreover, all the operators discussed in Section \ref{expr} are implemented in \setlog as a library called \Verb+arraylib.slog+.\footnote{This file is available here \urlarr.}

There are some minor syntactic differences between $\LQA$ and the actual language accepted by \setlog:
\begin{enumerate*}
\item In \setlog variables must begin with a capital letter because otherwise the identifier is interpreted as a constant;
\item The interval $[k,m]$ is written as \Verb+int(k,m)+;
\item The extensional set $\{x \plus E\}$ is written \Verb+{x / E}+;
\item The ordered pair $(x,y)$ is written $[x,y]$;
\item The nested RUQ $(\forall x \in E, y \in F: \phi)$ is written \Verb+foreach([X in E, Y in F],+$\phi$\Verb+)+; 
\item $\neq$ is written \Verb+neq+, $\in$ is written \Verb+in+ and $\notin$ is written \Verb+nin+; and
\item Conjunction is written with \Verb+&+ and implication is written \Verb+implies+.
\end{enumerate*}
Besides, the arguments of intervals can only be numbers and variables.
Then, for instance, $[k+1,m]$ (with $k$ and $m$ variables) is written as \Verb+int(J,M)+ plus the constraint \Verb.J is K + 1., where \Verb+J+ is a new variable and \Verb+is+ is the standard Prolog predicate that forces the evaluation of integer expressions.

\setlog can be used as an automated theorem prover but it can also be used as a CLP language.
In both cases it can be used as a Prolog library or interactively in a similar way to Prolog.
In the latter form, we can query \setlog interactively as we do in these simple examples using arrays.
\begin{Verbatim}
?- consult('setlog.pl').
?- setlog.
{log}=> add_lib('arraylib.slog').
{log}=> arr(A,5).
  A = {[1,_N5],[2,_N4],[3,_N3],[4,_N2],[5,_N1]}     % _N fresh variable
{log}=> arr(A,5) & foreach([X,Y] in A, Y = X).
  A = {[1,1],[2,2],[3,3],[4,4],[5,5]}
{log}=> arr(A,5) & arr(B,2) & un(A,B,C).
  A = {[1,_N7],[2,_N6],[3,_N5],[4,_N4],[5,_N3]},  
  B = {[1,_N2],[2,_N1]},  
  C = {[1,_N7],[2,_N6],[3,_N5],[4,_N4],[5,_N3],[1,_N2],[2,_N1]}
{log}=> arr(A,5) & arr(B,2) & un(A,B,C) & arr(C,N).
  A = {[1,_N5],[2,_N4],[3,_N3],[4,_N2],[5,_N1]},  
  B = {[1,_N5],[2,_N4]},  
  C = {[3,_N3],[4,_N2],[5,_N1],[1,_N5],[2,_N4]},  
  N = 5
{log}=> arr(A,5) & arr(B,2) & un(A,B,C) & arr(C,N) & 5 < N.
  no
{log}=> arr(A,N) & 0 < N & 0 < M & M < N & foreach([X,Y] in A, X neq M).
  no
\end{Verbatim} 
Note that the last two answers are \Verb+no+, meaning that those formulas are unsatisfiable.
The answers to the first queries are counterexamples disproving the unsatisfiability of the corresponding formulas.
In all cases variables of the form \Verb+_N+$\langle \mathit{number} \rangle$ are fresh variables.

It is important to remark that \setlog supports a language far beyond $\LQA$---see \setlog user's manual for a detailed user-oriented presentation \cite{Rossi00}.
If users want to use exactly the results shown in this paper they must take care of entering only $\LQA$ formulas.

\section{\label{applications}APPLICATIONS}
In this section we present two concrete applications of the implementation of $\SQA$ in \setlog aimed at formal verification and test case generation.
The 7 $\SQA$ formulas introduced below and other 36 verification conditions derived from the verification of 10 programs with arrays can be found here: \urlarr.
\setlog is able to solve all the 43 formulas in around 8 seconds on a standard laptop computer.

\subsection{Using \setlog as an automated theorem prover}
Being a satisfiability solver \setlog can be used to automatically discharge verification conditions laying in the implemented decidable fragments---e.g., \cite{DBLP:journals/jar/CristiaR21b,DBLP:journals/jar/CristiaLL23,DBLP:conf/csfw/CapozuccaCHK24}.
In this way we can use \setlog's implementation of $\SQA$ to discharge verification conditions (VC) of some classes of programs with arrays. As an example, consider the pseudocode of binary search shown in Figure \ref{f:bs}, where the gray boxes are program annotations indicating, respectively, the pre-condition, a loop invariant and the post-condition.

$\mathit{sorted}(A,n,k,m)$ is a predicate defined in the \Verb+arraylib.slog+ library stating that if $1 \leq k \leq m \leq n$ then array $A$ is sorted between components $k$ and $m$, where $n$ is supposed to be $A$'s length.
Observe that all the predicates in the annotations are $\LQA$ formulas.
The VC derived from the instructions and the annotations are shown in Figure \ref{f:vcbs}.
For example, the VC named \textsc{Invariance Lemma} is the result of:
\begin{enumerate*}
\item Assuming the pre-condition and the loop condition (line 2) plus the invariant at the beginning of the loop (lines 3 and 4);
\item Then the assertions corresponding to the loop body (line 5, corresponding to \Verb. m := (L + R) div 2.; and line 6, corresponding to the conditional instruction) are conjoined;
\item Finally, the invariant at the end of the loop should be implied (lines 7 and 8).
\end{enumerate*}
Note that the VC is enclosed in a \Verb+neg+ predicate which implements logical negation in \setlog.
This is so because \setlog should prove the unsatisfiability of the VC.
Also observe that variables \Verb+L_+ and \Verb+R_+ represent the values of \Verb+L+ and \Verb+R+ after the execution of the loop body.

\newcommand{\prebs}{\colorbox{gray!15}{\textsc{Pre}: $\Arr(A,N) \land \mathit{sorted}(A,N,1,N)$}}
\newcommand{\invbs}{\colorbox{gray!15}{\textsc{Inv}: $(\forall (i,y) \in A: 1 \leq i < L \implies y < x) \land (\forall (i,y) \in A: R \leq i \leq N \implies x \leq y)$}}
\newcommand{\postbs}{\colorbox{gray!15}{\textsc{Post}: $\Get(A,R,x) \lor (\forall (i,y) \in A: y \neq x)$}}

\begin{figure}
\begin{Verbatim}[commandchars=\\\{\}]
\prebs
int binarysearch(A,N,x)
begin
  L := 1; R := N;
\invbs
  while L < R do
    m := (L + R) div 2;
    if A[m] < x then L := m + 1 else R := m end
  end
  if A[R] = x then return R else return -1
end
\postbs
\end{Verbatim}
\caption{\label{f:bs}An implementation of binary search. Gray boxes are program annotations.}
\end{figure}

\begin{figure}
\begin{tabular}{c}
\toprule
\textsc{The invariant holds right before the loop} \\[1mm]
\begin{minipage}{\textwidth}
\begin{Verbatim}
foreach([I,Y] in A, 1 =< I & I < 1 implies Y < X) &
foreach([I,Y] in A, N =< I & I =< N implies X =< Y).
\end{Verbatim}
\end{minipage}
\\[1mm]\midrule
\textsc{Invariance Lemma: The loop's body preserves the invariant} \\
\begin{minipage}{\textwidth}
\begin{Verbatim}[numbers=left]
neg(
  arr(A,N) & sorted(A,N,1,N) &
  foreach([I,Y] in A, 1 =< I & I < L implies Y < X) &
  foreach([I,Y] in A, R =< I & I =< N implies X =< Y) &
  L + R =:= 2*M + H & 0 =< H & H =< 1 & get(A,M,Z) &
  (Z < X & L_ is M + 1 & R_ = R or neg(Z < X) & R_ is M & L_ = L)
  implies  foreach([I,Y] in A, 1 =< I & I < L_ implies Y < X) &
           foreach([I,Y] in A, R_ =< I & I =< N implies X =< Y)
).
\end{Verbatim}
\end{minipage}
\\[1mm]\midrule
\textsc{When the loop terminates the post-condition holds} \\
\begin{minipage}{\textwidth}
\begin{Verbatim}
neg(
  arr(A,N) & sorted(A,N,1,N) &
  R =< L &
  foreach([I,Y] in A, 1 =< I & I < L implies Y < X) &
  foreach([I,Y] in A, R =< I & I =< N implies X =< Y)
  implies  (get(A,R,X) or foreach([I,Y] in A, Y neq X))
).
\end{Verbatim}
\end{minipage}
\\\bottomrule
\end{tabular}
\caption{\label{f:vcbs}Verification conditions for the program of Figure \ref{f:bs}}
\end{figure}

\subsection{Using \setlog as a test case generator}
\setlog can also be used as a test case generator \cite{CristiaRossiSEFM13}.
After implementing $\SQA$, \setlog can generate test cases for programs with arrays by following a model-based testing method known as the Test Template Framework (TTF) \cite{DBLP:journals/tse/StocksC96} already implemented in \setlog \cite[Section 6]{CRISTIA_CAPOZUCCA_ROSSI_2026}.
In the TTF test cases are witnesses satisfying formulas specifying the conditions under which the system must be tested.

Consider the implementation of binary search shown above.
A good functional testing coverage starts by considering arrays of length 1 to 5.
Below we show some representative test specifications and cases for arrays of length 5.
In all cases \Verb+A+ is the array and \Verb+X+ is the sought element.
In order to generate test cases \setlog is asked to generate ground solutions. 
\newcommand{\testA}{\textrm{\small \Verb+X+ is greater than all the elements in \Verb+A+; \Verb+A+ is not a constant array but has some repeated elements}}
\newcommand{\testB}{\textrm{\small \Verb+X+ is not in \Verb+A+; there are elements in \Verb+A+ greater than and less than \Verb+X+; \Verb+A+ is an injective array}}
\newcommand{\testC}{\textrm{\small \Verb+X+ is in the ``middle'' of \Verb+A+ but is not the middle element; \Verb+A+ is an injective array}}
\newcommand{\testD}{\textrm{\small \Verb+X+ is the first element of \Verb+A+; \Verb+A+ is an injective array}}
\begin{Verbatim}[commandchars=\\\|\|]
{log}=> groundsol.
\testA
{log}=> arr(A,5) & sorted(A) & un(B,C,A) & size(B,3) & ipfun(B) &
        foreach([I,Y] in A, Y < X).
  A = {[1,-2],[2,-1],[3,0],[4,0],[5,0]},  X = 1
\testB
{log}=> arr(A,5) & sorted(A) & ipfun(A) & [_,V] in A & [_,W] in A &
        V < X & X < W & foreach([I,Y] in A, Y neq X).
  A = {[1,-1],[2,1],[3,2],[4,3],[5,4]},  X = 0
\testC
{log}=> arr(A,5) & sorted(A) & ipfun(A) & [_,V] in A & [_,W] in A &
        [I,X] in A & V < X & X < W & I neq 3.
  A = {[1,-1],[2,0],[3,1],[4,2],[5,3]},  X = 0
\testD
{log}=> arr(A,5) & sorted(A) & ipfun(A) & [1,X] in A.
  A = {[1,0],[2,1],[3,2],[4,3],[5,4]},  X = 0
\end{Verbatim}

\section{\label{relwork}DISCUSSION AND RELATED WORK}

After Bradley's work a few other researchers have proposed other approaches to the problem of the decidability of a theory of arrays.
Habermehl et al. \cite{DBLP:conf/fossacs/HabermehlIV08} propose a decidable logic for reasoning about infinite arrays of integers.
This logic is in the $\exists^*\forall^*$ fragment allowing for the expression of difference constraints and periodicity constraints on universally quantified indices as well as constraints on consecutive elements of arrays---e.g., $\forall i . 0 \leq i < n \rightarrow a[i+1] = a[i] - 1$.
Habermehl result is proved by building an automaton from a given formula in the array logic and then using known results about decidability over such automata. As can be seen Habermehl's logic is more expressive than Bradley's although it cannot express $\neq$ constraints involving array indexes, as in $\LQA$.
The same authors apply some of their results to the verification of integer array programs \cite{Bozga2009}.

de Moura and Bj{\o}rner \cite{DBLP:conf/fmcad/MouraB09} present   combinatory array logic (\textsf{CAL}) which permits the expression of complex properties over arrays.
\textsf{CAL} is defined on top of a core solver as a collection of inference rules.
In turn, the core solver is based on the standard architecture of SMT solvers. The core solver includes decision procedures for Boolean logic, arithmetic, bit-vectors and scalar values. The array theory present in \textsf{CAL} builds on the two basic operations, read and write. The latter is called an array combinator, because it builds new arrays. This basic combinator is extended with constant-value and map combinators which further extend the expressiveness of the logic. In fact, these combinators allows for the expression of some simple set theoretic operators such as union and intersection. \textsf{CAL} is implemented as part of the Z3 SMT solver.
More recently, Raya and Kun\u{c}ak \cite{DBLP:journals/jlap/RayaK24} study \textsf{CAL} from an algebraic perspective. In this way, they manage to assess the expressiveness of this fragment by analyzing the satisﬁability of array formulas as the satisﬁability of the theory of indexes and the theory of elements. As a result, \textsf{CAL} becomes a fragment of QFBAPAI which in turn is a fragment of RegExp-QFBAPA-Power.
They further extend the previous result in a way that some fragments discussed in the literature can be studied within the SymRegExp-QFBAPA-Power theory \cite{DBLP:journals/jlap/RayaK24a}.

Falke, Merz and Sinz \cite{Falke2013} extend the theory of arrays with $\lambda$-expressions that allow to express operations such as \verb+memset+ or \verb+memcpy+ and array initializations---i.e., loops where every array component is initialized. Satisfiability of quantifier-free formulas in this extension is proved to be decidable by encoding it in decidable theories available in SMT solvers. $\LQA$ allows to express some of these properties. For example, the array resulting after the execution of the following initialization:
\begin{Verbatim}
for(i = 0; i < n; ++i) {A[i] = i;}
\end{Verbatim}
can be expressed in $\LQA$ as $(\forall (x,y) \in A: y = x)$ assuming $\Arr(A,n)$ holds. However, $\LQA$ cannot express more complex initializations such as \Verb.A[i] = i + 1. that are expressible with $\lambda$-expressions.

Alberti et al. \cite{DBLP:journals/jar/AlbertiGS15} present two decidability results for flat array properties. Consider a quantifier-free, decidable theory $T$ extended with infinitely many free unary function symbols (i.e., the arrays). A flat array property in that extension of $T$ is a formula in the fragment $\exists^*\forall$ where for every term of the form $a[t]$, term $t$ is always a variable. 
These authors prove that if the $T$-satisfiability of formulas of the form $\exists^*\forall\exists^*$ is decidable, then the satisfiability of flat array properties is decidable. A second result is proved when the sorts of indexes and elements of arrays are different.
In spite that the quantifier-free theory defined in $\LQA$ is not as general as Alberti's, our languages does support the $\exists^*\forall$ fragment for array formulas.
In a later work \cite{DBLP:journals/fmsd/AlbertiGP17} the same authors augment the previous result with interpreted sets (i.e., set comprehensions) of a specific form. Then, they prove that some
fragments of this new logic are both decidable and expressive enough as to model
and reason about problems of fault-tolerant distributed systems. 

The Array Folds Logic (AFL), defined by Daca and others \cite{Daca2016}, is an extension of the quantifier-free theory of integer arrays allowing the expression of counting. Some properties expressible in AFL are not expressible in quantified fragments of the theory of arrays (e.g., Bradley's and Habermehl's) and, similarly, some properties (e.g., sortedness) expressible in the latter are not expressible in AFL.

Wang and Appel \cite{DBLP:journals/jar/WangA23} study the decidability of a theory of arrays with concatenation by extending Bradley's result.
Since this new fragment is undecidable the authors characterize a ``tangle-free'' fragment which happens to be decidable. A formula in the fragment with concatenation is tangle-free if there is no index shifting. In turn, index shifting results from the presence of terms of the form $a[i]$ and $a[i+n]$ ($n$ constant or variable) in the same quantified formula.
$\LQA$ cannot express concatenation as it cannot even express sums on array indexes.

A quite different approach w.r.t. those already discussed is followed by Plazar et al. \cite{Plazar2017}. This group applies Constraint Programming over Finite Domains to solve problems involving arrays. In previous works they manage to solve the case of arrays having a fixed, finite size. Then, they extend that result to the general case where the size of arrays is a variable by transforming a quantifier-free formula over those arrays into an equivalent formula with only fixed-size arrays, at which point the previous solvers can be called. The resulting technique competes in efficiency with existing SMT solvers.

The work by De Angelis et al. \cite{DBLP:journals/fuin/AngelisFPP17a} is somewhat related to ours as it is based on CLP although it is applied in a quite different way. They encode properties of imperative programs with arrays as CLP programs. The execution of these CLP programs is equivalent to (dis)proving the intended properties. Given that this method is not complete, when the answer of their method is inconclusive, the authors rest on SMT solvers to try to solve the problem.
As shown by their experiments, SMT solvers tend to be more effective when called after the CLP techniques where applied due to the propagation of the pre- and post-conditions performed by the transformation.

As can bee seen, our results are not only slightly more expressive than Bradley's but they also overlap with some of the results discussed above.
An important line of research departing from our present work is to study whether or not set theory can be used as a unifying theory where results such as Bradley's, Habermehl's and Alberti's, to name some of them, can be analyzed and proved.
For example, the $\lambda$-expressions defined by Falke can be put in terms of set theory by means of RUQ or restricted intensional sets \cite{DBLP:journals/jar/CristiaR21a}.
As another example, consider the kind of problems solved by Daca et al. For instance,  their problem ``given an array, accept it if the number of minimum elements in the array is the same as the number of maximum elements in the array'' is a $\LQA$-expressible problem. However, can $\LQA$ cover the full of array folds logic?
If not, can $\LQA$ be extended as to do so in a decidable manner?
Therefore, the general question is: Can set theory model all the results discussed in this section? Can these results be extended by taking advantage of decidability results available in set theory (e.g., in computable set theory, \cite{DBLP:series/mcs/CantoneOP01})? Would the algorithms obtained in this way be efficient enough?
It is worth to be mentioned that Raya and Kun\u{c}ak pose similar questions but their approach is different as is not explicitly based on set theory.

\section{\label{concl}FINAL REMARKS}

The main conclusion of the present work is that set theory can be used to model and reason about programs with arrays. 
This is a first result showing similar capabilities with respect to the seminal work by Bradley et al. \cite{DBLP:conf/vmcai/BradleyMS06}.
The theoretical results presented here have been implemented as part of \setlog, a tool based on set theory and constraint logic programming which now can be used to reason about programs with arrays.
The application of \setlog to program verification and test case generation has also been shown.

As we have said, this is a first result showing that set theory can be used as a logic to study the problem of analyzing programs with arrays. We believe set theory can be further used to reach results beyond Bradley's. This will be our first and foremost line of work. By following it we intend to answer some of the questions posed when we compared our work with the literature.

\bibliographystyle{ACM-Reference-Format}
\bibliography{/home/mcristia/escritos/biblio}

\end{document}